\documentclass[
 aps,
 ,twocolumn
,showpacs]{revtex4}
\usepackage{graphicx}
\usepackage{amsmath}
\usepackage{amssymb}
\usepackage{mathrsfs}
\usepackage{latexsym}
\usepackage{color}

\usepackage{tikz}
\usepackage{pgflibraryarrows}
\usepackage{pgflibrarysnakes}
\usetikzlibrary{decorations.pathmorphing}
\usetikzlibrary{patterns}

\usepackage{hyperref}

\begin{document}

\title{Black Hole Squeezers}

\author{Daiqin Su$^{1}$}
\email{sudaiqin@gmail.com}
\author{C. T. Marco Ho$^{1}$}
\author{Robert B. Mann$^{2,3}$}
\email{rbmann@uwaterloo.ca}
\author{Timothy C. Ralph$^{1}$}
\email{ralph@physics.uq.edu.au}
\affiliation{$^1$Centre for Quantum Computation and Communication Technology, School of Mathematics and Physics, The University
of Queensland, St. Lucia, Queensland, 4072, Australia}
\affiliation{$^2$Perimeter Institute, 31 Caroline Street North, Waterloo, Ontario N2L 2Y5, Canada}
\affiliation{$^3$ Department of Physics and Astronomy, University of Waterloo, Ontario, Canada, N2L 3G1}


\date{\today}


\begin{abstract}
We show that  the gravitational quasi-normal modes (QNMs) of a Schwarzschild black hole play 
the role of a multimode squeezer that can generate particles. For a minimally coupled scalar field, the QNMs
``squeeze" the initial state of the scalar field (even for the vacuum) and produce scalar particles.
The maximal squeezing amplitude is inversely proportional to the cube of the imaginary part of the QNM
frequency, implying that the particle generation efficiency is higher for lower decaying QNMs. Our results show that the gravitational 
perturbations can amplify Hawking radiation.   
\end{abstract}

\maketitle

\vspace{10 mm}

\section{Introduction}

Particle creation  is a fundamental phenomenon characteristic of 
quantum field theory in curved spacetime \cite{Birrell84, Wald94, Parker09}. It typically occurs in highly dynamical spacetimes:  in the very early universe,  initial quantum fluctuations of curvature can be amplified by the exponentially expanding universe to 
form tiny perturbations on the background spacetime \cite{Sakharov65, Mukhanov92}. These
 play a crucial role in explaining the anisotropies of the cosmic microwave background (CMB) and the formation of the large scale structure of the observed universe. Primordial gravitational waves are also expected to be generated in the early universe and 
their detection is one of the main targets of modern astrophysics. In addition to dynamical
spacetimes, stationary or static spacetimes can also create quantum particles. Well-known examples include the superradiance from a rotating black hole \cite{Zeldovich72, Misner72, Unruh74}, the Unruh-Davies \cite{Unruh76, Davies75} radiation observed by a uniformly accelerated observer, 
and the Hawking radiation from a Schwarzschild black hole \cite{Hawking75}. Although Hawking
particles were observed in an analogue system recently \cite{Steinhauer16}, detection of  Hawking radiation from a real black hole remains 
elusive because the temperature of a solar mass black hole ($~ 10^{-8}$ $^o$K) is much lower than the temperature of the CMB ($\sim 2.7$  $^o$K).  

One question  of particular interest is whether gravitational waves (ripples of  spacetime) can create quantum particles. Although a dynamical spacetime characteristically generates particles, 
 it has been shown that particle creation by plane gravitational waves is forbidden \cite{Gibbons75, Deser75, Garriga91, Sorge00}. A similar statement applies to electromagnetic waves: electron-positron pairs cannot be produced by plane electromagnetic waves, no matter how strong we make the electromagnetic  field \cite{Schwinger51}; otherwise 
momentum conservation would be violated. However pair production of electrons and positrons is possible if a nucleus is introduced to balance the momentum \cite{Hubbell06}. 

In the gravitational wave case, one might expect that an analog to the nucleus, e.g., a black hole, has to be introduced to allow particle creation. 
From a theoretical perspective, a detailed study of this issue is important, and necessary 
to determine whether and to what degree gravitational 
perturbations in a black hole spacetime can create particles. There are now three gravitational wave events that have originated from the coalescence of two black holes that have been directly detected by the
Laser Interferometer Gravitational-Wave Observatory (LIGO) \cite{GW16:1,GW16:2,Abbott:2017vtc}. 
The observed gravitational wave signals reveals that the spacetime changes dramatically
when  two black holes merge into one, with a large amount of energy of order a few solar rest masses  carried away by the emitted gravitational waves. If particles, e.g., photons, can be produced by  gravitational perturbations they will travel
along with the gravitational waves and could be detected if the particle creation efficiency is high enough. From an observational perspective, it is therefore also very important to have a thorough study of this problem.
  
In this paper we address this question.  Instead of studying the whole process of the coalescence of two black holes, 
a very complicated situation requiring numerical relativity \cite{Lehner01},
we study the final stage of merging: the ring-down stage. At the ring-down stage, the quasi-normal oscillations of the black hole are dominant, which are known as 
the gravitational quasi-normal modes (QNMs). We are interested in the effects of the gravitational quasi-normal
modes (QNMs) of a black hole, which have been extensively studied for decades \cite{Price72,Chan:1996yk,Nollert99, Kokkotas99, Berti09, Konoplya11} and to which analytic techniques can be applied. We consider a massless Hermitian scalar field
that propagates in the Schwarzschild background spacetime with quasi-normal perturbations. The scalar field is assumed to be minimally coupled with the spacetime. 
The coupling can be divided into two parts:  with the Schwarzschild background spacetime and with the QNMs. Though the former is well studied, the latter is somewhat less understood and is the main content of this paper. 

We derive the interaction Hamiltonian for the scalar field, which implies that
the QNMs play the role of a multimode squeezer. We show that the QNMs ``squeeze" the initial state (vacuum or thermal state) of 
the scalar field and produce particles. In this sense black holes themselves can be quantum squeezers.

Our paper is organized as follows. In Sec. \ref{SF}, we briefly review the quantization of a massless scalar field in the Schwarzschild background 
spacetime. In Sec. \ref{QNM}, we review the gravitational QNMs for Schwarzschild black holes and list some important
results for our calculations. In Sec. \ref{Coupling}, we study the coupling between the scalar field and the gravitational QNMs, and
derive the interaction Hamiltonian for the scalar field, based on which we show that the gravitational perturbations around a 
Schwarzschild black hole create particles. In Sec. \ref{PNEstimation}, we provide a crude estimate of the total particle number. 
We summarize with some concluding remarks in Sec. \ref{conclusion}. Throughout this paper we work with units $\hbar =  G = c =1$.

\section{Scalar field in curved spacetime}\label{SF}

We consider a Hermitian massless scalar field $\Phi$ that minimally couples to the curved spacetime with metric $g_{\mu\nu}$. 
The Lagrangian density for the scalar field is \cite{Birrell84}
\begin{equation}\label{Lagrangian}
\mathcal{L} = \frac{1}{2} \sqrt{-g} g^{\mu\nu}(\partial_{\mu} \Phi) (\partial_{\nu} \Phi) ,
\end{equation}
where $g$ is the determinant of $g_{\mu\nu}$. We assume that the metric $g_{\mu\nu}$ can be decomposed into a
background part $g_{B\mu\nu}$ and a perturbation  $h_{\mu\nu}$, namely, $g_{\mu\nu} = g_{B\mu\nu} + h_{\mu\nu}$. 
The background metric usually possesses some symmetries (time-translation invariance, rotational invariance etc.) and the dynamics of the scalar
field in the background spacetime is well established. The perturbation $h_{\mu\nu}$ is assumed to be small so that perturbation theory is applicable. 
Expanding the Lagrangian density Eq. (\ref{Lagrangian}) with respect to $h_{\mu\nu}$ and keeping terms to first order, we find
\begin{eqnarray}
\mathcal{L} &=& \mathcal{L}_0 + \mathcal{L}_1, 
\end{eqnarray}
where the background part $\mathcal{L}_0$ and perturbed part $\mathcal{L}_1$ are
\begin{eqnarray*}
\mathcal{L}_0 &=& \frac{1}{2} \sqrt{-g_B} g_B^{\mu\nu}(\partial_{\mu} \Phi) (\partial_{\nu} \Phi) , \nonumber\\
\mathcal{L}_1 &=& \frac{1}{4} \sqrt{-g_B} \big( h^{\alpha}_{~\alpha} g_B^{\mu\nu} - 2 h^{\mu\nu} \big) (\partial_{\mu} \Phi) (\partial_{\nu} \Phi),
\end{eqnarray*}
with $g_B$ the determinant of the background metric and $ h^{\alpha}_{~\alpha} \equiv g_{B\alpha\beta}h^{\alpha\beta}$ the trace of the metric perturbation. 
Note that we use the convention: $h^{\mu \nu} \equiv g_B^{\mu \alpha} g_B^{\nu \beta} h_{\alpha \beta}$. 
In this paper, we are concerned with the Schwarzschild background spacetime, for which the line element in the Schwarzschild coordinates $(t, r, \theta, \phi)$ is
\begin{equation}\label{Schwarzschild}
\mathrm d s^2 = - f(r) \mathrm d t^2 + \frac{1}{f(r)} \mathrm d r^2 + r^2( \mathrm d \theta^2 + \sin^2 \theta \mathrm d \phi^2),
\end{equation}
where $f(r) = 1 - 2M/r$ and $M$ is the mass of the Schwarzschild black hole. 

The canonically conjugate field of $\Phi$ is also decomposed into a background and perturbed part, 
\begin{eqnarray}
\Pi &=& 
\Pi_0 + \Pi_1, 
\end{eqnarray}
where 
\begin{eqnarray*}
\Pi_0 &=& \frac{\partial \mathcal{L}_0}{\partial(\partial_t \Phi)} = \sqrt{- g_B} g_B^{tt} (\partial_t \Phi), \nonumber\\
\Pi_1 &=& \frac{\partial \mathcal{L}_1}{\partial(\partial_t \Phi)} = - \frac{1}{2} \sqrt{- g_B} \big[ 2 h^{t\nu} (\partial_{\nu} \Phi) - h^{\alpha}_{~\alpha} g_B^{tt} (\partial_t \Phi) \big].
\end{eqnarray*}
The Hamiltonian density is 
\begin{equation}
\mathcal{H} = 
\mathcal{H}_0 + \mathcal{H}_1,
\end{equation}
where 
\begin{eqnarray}
\mathcal{H}_0 &=&  \Pi_0 (\partial_t \Phi) - \mathcal{L}_0 \nonumber \\
&=& \frac{1}{2} \sqrt{-g_B} \big[ g_B^{tt}(\partial_t \Phi)^2 - g_B^{i j}(\partial_i \Phi) (\partial_j \Phi) \big]
\end{eqnarray}
is the unperturbed Hamiltonian density and 
\begin{eqnarray}
\mathcal{H}_1 &=& \Pi_1 (\partial_t \Phi) - \mathcal{L}_1 \nonumber \\
&=&-\frac{1}{2} \sqrt{-g_B} \big[h^{tt} (\partial_t \Phi)^2 - h^{ij}(\partial_i \Phi) (\partial_j \Phi)\big]
+ \frac{1}{2} h^{\alpha}_{~\alpha}\mathcal{H}_0  \nonumber\\
\end{eqnarray}
is the perturbed Hamiltonian density.
For the Schwarzschild background spacetime, 
$\sqrt{-g_B} = r^2 \sin \theta$, so the perturbed Hamiltonian is
\begin{eqnarray}\label{pHamiltonian}
H_1 &=& \int \mathrm d^3x \mathcal{H}_1 \nonumber\\
&=& \frac{1}{2} \int_{2M}^{\infty} \mathrm d r \int_{4 \pi} \mathrm d \Omega ~ r^2 \bigg\{-h^{tt} (\partial_t \Phi)^2 + h^{ij}(\partial_i \Phi) (\partial_j \Phi) \nonumber\\
&&+ \frac{1}{2} h^{\alpha}_{~\alpha}\big[ g_B^{tt}(\partial_t \Phi)^2 - g_B^{i j}(\partial_i \Phi) (\partial_j \Phi) \big] \bigg\}
\end{eqnarray}
where $\mathrm d \Omega = \sin \theta \mathrm d \theta \mathrm d \phi$. 

The dynamics of the scalar field on the background spacetime is determined by the unperturbed Lagrangian density $\mathcal{L}_0$, from which
one can derive the Klein-Gordon equation \cite{Birrell84},
\begin{equation}\label{fieldequation}
\frac{1}{\sqrt{-g_B}} \partial_{\mu} \big(\sqrt{-g_B}g_B^{\mu\nu} \partial_{\nu} \Phi \big) = 0. 
\end{equation}
The normal-mode solutions to Eq. (\ref{fieldequation}) can be decomposed as 
\begin{equation}
\phi_{\omega l m}(t, r, \theta, \phi) = \frac{1}{\sqrt{4\pi \omega}} e^{-i\omega t} Y_{lm} (\theta, \phi) R_{\omega l}(r)/r
\end{equation}
where $\omega>0$ is the frequency of the mode, $Y_{lm} (\theta, \phi)$ is the spherical harmonic. The radial function $R_{\omega l}(r)$ satisfies
\begin{equation}
- \frac{d^2 R_{\omega l}}{d r_*^2} + V^{(s)}_l(r) R_{\omega l} = \omega^2 R_{\omega l},
\end{equation}
where $V^{(s)}_l(r)$ is the effective potential
\begin{equation}\label{EffPot:Scalar}
V^{(s)}_l(r) = f(r) \bigg( \frac{l(l+1)}{r^2} + \frac{2M}{r^3} \bigg).
\end{equation}
Here $r_*$ is the tortoise coordinate
\begin{equation}
d r_* = d r/f(r), ~~~~~~ r_* = r + 2M \ln (r/2M - 1). 
\end{equation}
Note the $r_* \rightarrow -\infty$ corresponds to the event horizon of the Schwarzschild black hole.
An implicit relation $r=r(r_*)$ can be derived and substituted into the effective potential $V^{(s)}_l$, Eq. (\ref{EffPot:Scalar}), so that the effective potential can be considered as a function of $r_*$. 

We only consider the field outside the event horizon, where there exist two sets of orthonormal modes that can completely represent the field.
They are the upcoming and ingoing modes, denoted as $\phi^{\text{up}}_{\omega l m}$ and  $\phi^{\text{in}}_{\omega l m}$, respectively. 
The asymptotic behaviour for the radial part of the upcoming mode, $R^{\text{up}}_{\omega l}$, is
\begin{eqnarray}\label{upasymptotic}
R^{\text{up}}_{\omega l} \sim 
\begin{cases}
B^{\text{up}}_{\omega l} e^{i\omega r_*}, & r_* \rightarrow +\infty;  \\
 e^{i\omega r_*} + A^{\text{up}}_{\omega l} e^{-i\omega r_*} , &  r_* \rightarrow -\infty, 
\end{cases}
\end{eqnarray}\label{inasymptotic}
and for the radial part of the ingoing mode,  $R^{\text{in}}_{\omega l}$, is
\begin{eqnarray}
R^{\text{in}}_{\omega l} \sim
\begin{cases}
e^{-i\omega r_*} + A^{\text{in}}_{\omega l} e^{i\omega r_*}, &  r_* \rightarrow +\infty; \\
B^{\text{in}}_{\omega l} e^{-i\omega r_*}, &  r_* \rightarrow -\infty.
\end{cases}
\end{eqnarray}
Here $A^{\text{up}}_{\omega l}$ ($A^{\text{in}}_{\omega l}$) and $B^{\text{up}}_{\omega l}$ ($B^{\text{in}}_{\omega l}$) are the reflection and transmission amplitudes of the upcoming (ingoing) modes, respectively. They satisfy the following Wronskian relations \cite{HLO14},
\begin{eqnarray}
|A^{\text{up}}_{\omega l}|^2 &=& 1 - |B^{\text{up}}_{\omega l}|^2, \nonumber\\
|A^{\text{in}}_{\omega l}|^2 &=& 1 - |B^{\text{in}}_{\omega l}|^2, \nonumber\\
|A^{\text{up}}_{\omega l}| &=& |A^{\text{in}}_{\omega l}|, ~~~~
B^{\text{up}}_{\omega l} = B^{\text{in}}_{\omega l}.
\end{eqnarray}
The upcoming modes $\phi^{\text{up}}_{\omega l m}$ and ingoing modes $\phi^{\text{in}}_{\omega l m}$ are chosen to satisfy the orthonormality relations,
\begin{eqnarray}
\langle \phi^{\text{up}}_{\omega l m}, \phi^{\text{up}}_{\omega' l' m'} \rangle &=& \delta ({\omega-\omega'}) \delta_{ll'} \delta_{mm'} \nonumber \\
\langle \phi^{\text{up}*}_{\omega l m}, \phi^{\text{up}*}_{\omega' l' m'} \rangle &=& -\delta ({\omega-\omega'}) \delta_{ll'} \delta_{mm'} \nonumber \\
\langle \phi^{\text{in}}_{\omega l m}, \phi^{\text{in}}_{\omega' l' m'} \rangle &=& \delta ({\omega-\omega'}) \delta_{ll'} \delta_{mm'} \nonumber\\
\langle \phi^{\text{in}*}_{\omega l m}, \phi^{\text{in}*}_{\omega' l' m'} \rangle &=& -\delta ({\omega-\omega'}) \delta_{ll'} \delta_{mm'}, \nonumber\\
\langle \phi^{\text{up}}_{\omega l m}, \phi^{\text{in}}_{\omega' l' m'} \rangle &=& 0, \nonumber\\
\langle \phi^{\text{up}*}_{\omega l m}, \phi^{\text{in}*}_{\omega' l' m'} \rangle &=& 0. 
\end{eqnarray}
Here $\langle ~ , ~ \rangle$ represents the Klein-Gordon inner product \cite{Birrell84}, which is defined on a spacelike hypersurface $t=\text{const.}$ as
\begin{eqnarray}
\langle \varphi, \chi \rangle &=& i \int_{2M}^{\infty} \mathrm d r \frac{r^2}{f(r)} \int_{4\pi} \mathrm d \Omega 
\big(\varphi^* \partial_t \chi - \chi \partial_t \varphi^* \big)
\end{eqnarray}
for any two solutions $\varphi$ and $\chi$ of the Klein-Gordon equation (\ref{fieldequation}).

In the canonical quantization procedure, the scalar field $\Phi$ is regarded as an operator, satisfying certain canonical commutation relations, 
and is expanded as 
\begin{equation}\label{ScalarField}
\Phi = \sum^{\infty}_{l=0} \sum_{m=-l}^l \int_0^{\infty} \mathrm d \omega \bigg(\hat{a}_{\omega lm} \phi^{\text{up}}_{\omega lm} 
+ \hat{b}_{\omega lm} \phi^{\text{in}}_{\omega lm} + \text{h.c.} 
\bigg),
\end{equation}
where $\text{h.c.}$ represents the Hermitian conjugate. 
The operators $\hat{a}_{\omega lm}$ and $\hat{b}_{\omega lm}$ represent upcoming and ingoing modes, respectively. They satisfy the boson commutation relations
\begin{eqnarray}
&[\hat{a}_{\omega lm}, \hat{a}^{\dag}_{\omega' l'm'}] = \delta ({\omega-\omega'}) \delta_{ll'} \delta_{mm'}, ~\nonumber\\
&[\hat{b}_{\omega lm}, \hat{b}^{\dag}_{\omega' l'm'}] = \delta ({\omega-\omega'}) \delta_{ll'} \delta_{mm'}, \nonumber\\
&[\hat{a}_{\omega lm}, \hat{b}_{\omega' l'm'}] = [\hat{a}_{\omega lm}, \hat{b}^{\dag}_{\omega' l'm'}] = 0.
\end{eqnarray}

\section{Gravitational quasi-normal modes} \label{QNM}

A Schwarzschild black hole is a static and spherically symmetric spacetime that is described by the 
Schwarzschild metric Eq. (\ref{Schwarzschild}). Taking this metric to be the background metric ${g}_{B\mu\nu}$, 
gravitational perturbations $h_{\mu\nu} ={g}_{\mu\nu}- {g}_{B\mu\nu} $ can arise through various physical processes,
such as a star falling into the black hole. The equations governing the evolution of the perturbations were first derived
by Regge and Wheeler \cite{RW57}, and Zerilli \cite{Zerilli70}, in what is known as the Regge-Wheeler-Zerilli (RWZ) gauge. Due to the 
time-translation and rotational invariance of the Schwarzschild metric, the perturbations can be 
decomposed into eigenmodes with definite frequency and angular momentum. Furthermore, they can be classified as two distinct types:  
odd-parity (or magnetic-parity) and even-parity (or electric-parity) perturbations.  

In the RWZ gauge, the odd-parity perturbations are characterized by two functions $\tilde h_0(r)$ and $\tilde h_1(r)$. The nonzero 
components of $h_{\mu\nu}$ are
\begin{eqnarray}\label{MPO:RWZ}
h_{tA}^{(o)} &=& \tilde h_0(r) e^{-i\omega t} X^{lm}_A(\theta, \phi), \nonumber\\
h_{rA}^{(o)} &=& \tilde h_1(r) e^{-i\omega t} X^{lm}_A(\theta, \phi), 
\end{eqnarray}
where $A = \{ \theta, \phi\}$. Here $\omega$ is the frequency of the perturbations, and $X^{lm}_A$ is the odd-parity
vector spherical harmonic on the unit two-sphere \cite{Martel05},
\begin{equation}
X^{lm}_{\theta} = - \csc \theta ~ Y^{lm}_{~~~, \phi}, ~~~~~~ X^{lm}_{\phi} =  \sin \theta ~ Y^{lm}_{~~~, \theta},
\end{equation}
where $Y^{lm}(\theta, \phi)$ is the scalar spherical harmonic. The two functions $\tilde h_0(r)$ and $\tilde h_1(r)$ are not independent
and can be expressed in terms of a single scalar function $Q(r)$ as \cite{RW57}
\begin{equation}
\tilde h_0 = - \frac{f}{i \omega} \frac{d}{d r} (r Q), ~~~~ \tilde h_1 = \frac{r Q}{f}.
\end{equation}
The scalar function $Q(r)$ satisfies the equation
\begin{equation}\label{oddRW}
- \frac{d^2Q}{d r_*^2} + V^{(o)}_l Q = \omega^2 Q
\end{equation}
where
\begin{equation}
V^{(o)}_l(r) = f(r) \bigg(\frac{l(l+1)}{r^2} - \frac{6M}{r^3} \bigg)
\end{equation}
is the odd-parity effective potential.

In the RWZ gauge, the even-parity perturbations are characterized by three functions: $\tilde H_0(r), \tilde H_1(r)$ and $\tilde K(r)$.
The nonzero components of $h_{\mu\nu}$ are
\begin{eqnarray}\label{MPE:RWZ}
h_{tt}^{(e)} &=& f(r) \tilde H_0(r) e^{-i \omega t} Y^{lm}, ~~~~ h_{rr}^{(e)} = \frac{\tilde H_0(r)}{f(r)} e^{-i \omega t} Y^{lm}, \nonumber\\
h_{tr}^{(e)} &=& \tilde H_1(r) e^{-i \omega t} Y^{lm}, ~~~~ h_{AB}^{(e)} = r^2 \Omega_{AB} \tilde K(r) e^{-i \omega t} Y^{lm}, \nonumber\\
\end{eqnarray}
where $\Omega_{AB} = \text{diag}\{1, \sin^2 \theta \}$ is the metric on the unit two-sphere. $\tilde H_0(r), \tilde H_1(r)$ and $\tilde K(r)$ can be
expressed in terms of the Zerilli function $Z(r)$ as \cite{Nichols12}
\begin{eqnarray}
\tilde K &=& \bigg[\frac{\lambda(\lambda+1) r^2 + 3\lambda M r + 6 M^2}{r^2(\lambda r + 3 M)} \bigg] Z + \sqrt{f} Z_{,r}, \nonumber\\
\tilde H_1 &=& -i \omega \bigg[\frac{\lambda r^2 - 3\lambda M r - 3 M^2}{(r-2M)(\lambda r + 3 M)} \bigg] Z - i \omega r Z_{,r}, \nonumber\\
\tilde H_0 &=& \bigg[\frac{\lambda r (r-2M) - \omega^2 r^4 + M(r - 3M)}{(r-2M)(\lambda r + 3 M)} \bigg] \tilde K \nonumber \\
&& + \bigg[\frac{(\lambda+1) M - \omega^2 r^3 }{i \omega r (\lambda r + 3 M)} \bigg] \tilde H_1,
\end{eqnarray}
where
\begin{equation}
\lambda = \frac{1}{2}(l-1)(l+2).
\end{equation}
The Zerilli function satisfies the equation
\begin{equation}\label{evenRW}
- \frac{d^2Z}{d r_*^2} + V^{(e)}_l Z = \omega^2 Z
\end{equation}
with the even-parity effective potential
\begin{eqnarray}
&&V^{(e)}_l(r)  \nonumber \\
&=& f(r) \bigg[\frac{2\lambda^2(\lambda+1) r^3 + 6\lambda^2 M r^2 + 18 \lambda M^2 r + 18M^3}{r^3(\lambda r + 3 M)^2} \bigg]. \nonumber\\
\end{eqnarray}

The boundary conditions for the QNMs are that on the event horizon there is only an ingoing wave,
\begin{equation}
Q (Z) \sim e^{- i \omega r_*}, ~~~~ r_* \rightarrow - \infty ~( r \rightarrow 2M)
\end{equation}
and at spatial infinity there is only an outgoing wave, 
\begin{equation}
Q (Z) \sim e^{ i \omega r_*}, ~~~~ r_* \rightarrow + \infty ~( r \rightarrow + \infty). 
\end{equation}
The above boundary conditions imply that the perturbations are dissipative: waves can escape either to infinity or into the black hole.
The frequencies of the QNMs are complex, 
\begin{equation}
\omega = \omega_R - i \omega_I
\end{equation}
where $\omega_I$ is positive and characterizes the decay of the QNMs. For a Schwarzschild black hole, there is a discrete infinity of QNMs.
The QNM frequencies depend on $l$ and an integer $n$ called the overtone number \cite{Schutz85, Leaver85}.

Based on the Newman-Penrose (NP) null-tetrad formalism \cite{NP62}, another approach has been developed to study the gravitational perturbations in a 
Schwarzschild background \cite{Price72, Bardeen73}. This  more general method has been generalized to study the neutrino,
electromagnetic and gravitational perturbations in a Kerr background spacetime \cite{Teukolsky72, Teukolsky73, Press73, Teukolsky74}. In this framework,
 gravitational perturbations are represented by two field quantities $\psi_s$ with $s = \pm 2$, 
which are related to the Weyl scalars \cite{NP62} and satisfy the  Teukolsky master equation \cite{Teukolsky72}. 
For the Schwarzschild case, one can take the limit  $a\to 0$ ($a$ is the angular momentum per unit mass
of the Kerr black hole) in the master equation to obtain the corresponding field equation for $\psi_s$. This quantity  can be decomposed as
\begin{equation}
\psi_s = e^{-i \omega t} ~_sY_{lm}(\theta, \phi) ~_s R_{\omega l}(r)
\end{equation}
where $_sY_{lm}$ is the spin-weighted spherical harmonic \cite{Goldberg67}, $_s R_{\omega l}(r)$ is the radial function satisfying the equation
\begin{eqnarray}\label{TeukolskyEquation}
&\Delta_r^{-s} \frac{d}{dr}\bigg(\Delta_r^{s+1} \frac{d}{dr} \bigg) {}_s R_{\omega l}(r)
+ \bigg[ \frac{r^4 \omega^2 - 2 i s r^2 (r - M) \omega}{\Delta_r} \nonumber \\
&+ 4 i s \omega r - (l-s)(l+s+1) \bigg] {}_s R_{\omega l}(r) = 0
\end{eqnarray}
where $\Delta_r = r(r-2M)$. At large distance from the black hole, the asymptotic solutions of ${}_s R_{\omega l}$ are
\begin{equation}\label{QNinfinity}
{}_s R_{\omega l} \sim \frac{e^{-i \omega r_*}}{r}, ~~ \text{and}~~~ {}_s R_{\omega l} \sim \frac{e^{i \omega r_*}}{r^{2s+1}}
\end{equation}
whereas very close to the event horizon
\begin{equation}\label{QNhorizon}
{}_s R_{\omega l} \sim \Delta_r^{-s} e^{-i \omega r_*}, ~~ \text{and}~~~ {}_s R_{\omega l} \sim e^{i \omega r_*}. 
\end{equation}
The QNM boundary conditions  (only outgoing waves at spatial infinity, only
ingoing waves at the future horizon) imply
\begin{eqnarray}\label{boundary:QNM}
{}_s R_{\omega l} \sim 
\begin{cases}
C_{\omega l} ~ \Delta_r^{-s} e^{-i \omega r_*}, &  r_* \rightarrow -\infty; \\
D_{\omega l} ~ e^{i \omega r_*}/r^{2s+1}, & r_* \rightarrow +\infty. 
\end{cases}
\end{eqnarray}
where $C_{\omega l}$ and $D_{\omega l}$ are the amplitude of the QNM at the event horizon and spatial infinity, respectively.

The explicit expressions for the components of the metric perturbation $h_{\mu\nu}$ 
are very important when considering the coupling between the gravitational perturbations and the quantum fields. 
Chrzanowski, Cohen and Kegeles (CCK) developed a procedure for reconstructing the metric perturbation 
$h_{\mu\nu}$ in the {\it ingoing} and {\it outgoing} radiation gauges from the field quantity $\psi_s$ \cite{Chrzanowski75, Cohen74}.
Roughly speaking, the CCK procedure consists of two steps: The first step is to relate the field quantity $\psi_s$ to the so-called
Hertz potential $\Psi$, which also satisfies the master equation with spin weight $s=-2$; the second step is to find the 
relation between $h_{\mu\nu}$ and the Hertz potential $\Psi$ \cite{Chrzanowski75}. The first explicit calculation of 
the relation between the Hertz potential $\Psi$ and $\psi_s$ for the Schwarzschild black hole was done by Lousto and 
Whiting \cite{Lousto02}. Generalization to the Kerr black hole was performed by Ori \cite{Ori03}, Yunes and Gonzalez \cite{Yunes06}.

In the {\it ingoing} and {\it outgoing} radiation gauges, the trace of the metric perturbation, $h^{\alpha}_{~\alpha}$, vanishes in the whole spacetime \cite{Chrzanowski75}. 
The perturbation $h_{\mu\nu}$ is transverse at the past null infinity and at the future horizon in the {\it ingoing} radiation gauge. Therefore it is a suitable gauge to study the gravitational
effects near the event horizon. While in the {\it outgoing} radiation gauge, $h_{\mu\nu}$ is transverse at future null infinity and at the past horizon.  It is therefore a suitable gauge for studying gravitational effects at spatial infinity, e.g., gravitational waves emitted by a black hole. Since it is reasonable to expect that 
the interaction between the metric perturbation and the quantum fields is strong near the event horizon, we therefore work in the {\it ingoing} radiation  gauge throughout this paper. 

Following the CCK procedure, Nichols {\it et al} \cite{Nichols12} derived   explicit expressions for the metric perturbation  $h_{\mu\nu}$ in the Schwarzschild background spacetime
in the {\it ingoing} radiation gauge. For the odd (magnetic)-parity perturbations, the nonzero components are
\begin{widetext}
\begin{eqnarray}\label{OddPerturbation}
h_{t A}^{(o)} &=& - f h_{r A}^{(o)} =  \frac{\sqrt{D}}{2 f \sqrt{2l(l+1)}} \Re \bigg\{ \bigg[\frac{d}{d r_*}{}_{-2}R_{\omega l} - \bigg(i\omega + \frac{2 f}{r}  \bigg){}_{-2}R_{\omega l}\bigg] 
\big({}_{-1}Y_{lm} m_A + {}_{1}Y_{lm} m_A^* \big) e^{-i\omega t}\bigg\}, \nonumber\\
h_{AB}^{(o)} &=&\frac{1}{f^2} \Re \bigg\{ \bigg[(i \omega r^2 - M)\frac{d}{d r_*}{}_{-2}R_{\omega l} - \bigg(\frac{1}{2} \mu^2 f - i\omega (-3r+7M) - \omega^2 r^2  \bigg)
{}_{-2}R_{\omega l}\bigg] \big({}_{-2}Y_{lm} m_A  m_B - {}_{2}Y_{lm} m_A^* m_B^* \big) e^{-i\omega t}\bigg\}, \nonumber\\
\end{eqnarray}
where $D = (l+2)!/(l-2)!$, $\mu^2 = (l-1)(l+2)$,  $\Re$ represents the real part of a function, and
$m_A = \frac{1}{\sqrt{2}}(1, i \sin \theta)$ is a vector on the unit-sphere with its index raised by the metric $\Omega^{AB}$.
For the even (electric)-parity perturbations, the nonzero components are
\begin{eqnarray}\label{EvenPerturbation}
h_{tt}^{(e)} &=& - f h_{tr}^{(e)} = f^2 h_{rr}^{(e)}=-\frac{2\sqrt{D}}{r^2} \Re \big\{{}_{-2}R_{\omega l} ~ Y_{lm} e^{-i\omega t}\big\}, \nonumber\\
h_{t A}^{(e)} &=& - f h_{r A}^{(e)} = \frac{\sqrt{D}}{2 f \sqrt{2l(l+1)}} \Re \bigg\{ \bigg[\frac{d}{d r_*}{}_{-2}R_{\omega l} - \bigg(i\omega + \frac{2 f}{r}  \bigg){}_{-2}R_{\omega l}\bigg] 
\big({}_{-1}Y_{lm} m_A - {}_{1}Y_{lm} m_A^* \big) e^{-i\omega t}\bigg\}, \nonumber\\
h_{AB}^{(e)} &=&\frac{1}{f^2} \Re \bigg\{ \bigg[(i \omega r^2 - M)\frac{d}{d r_*}{}_{-2}R_{\omega l} - \bigg(\frac{1}{2} \mu^2 f - i\omega (-3r+7M) - \omega^2 r^2  \bigg)
{}_{-2}R_{\omega l}\bigg] \big({}_{-2}Y_{lm} m_A  m_B + {}_{2}Y_{lm} m_A^* m_B^* \big) e^{-i\omega t}\bigg\}. \nonumber\\
\end{eqnarray}
Note that the metric perturbation in the {\it ingoing} radiation gauge is related to that in the RWZ gauge, Eqs. (\ref{MPO:RWZ}) and (\ref{MPE:RWZ}), by a gauge transformation.
\end{widetext}

\section{Coupling between QNMs and scalar field}\label{Coupling}

In the absence of gravitational perturbations, the scalar field $\Phi$ evolves freely on the Schwarzschild background spacetime.
Its dynamics is dominated by the unperturbed Lagrangian density $\mathcal L_0$. If the Schwarzschild background spacetime is perturbed,
the scalar field will couple to the gravitational perturbations.  Its dynamics are governed by the interaction Hamiltonian
$H_1$ Eq. (\ref{pHamiltonian}). Our particular interest is in whether or not gravitational 
perturbations in a Schwarzschild background can produce particles.  We have noted earlier that  plane gravitational waves do not produce 
particles \cite{Gibbons75, Deser75, Garriga91, Sorge00}; were it otherwise, momentum conservation would be violated. 
As we shall demonstrate, the situation is different for spherical perturbations.   We will show  that gravitational perturbations in a 
Schwarzschild background do generate scalar particles and that angular momentum is conserved in this process.

In order to know the evolution of the state of the scalar field, one needs to find the explicit expression for the interaction
Hamiltonian $H_1$ which contains only first order terms of the components of the metric perturbation $h_{\mu\nu}$. An appropriate
gauge can be chosen so that the interaction Hamiltonian $H_1$ takes a relatively simple form. Throughout this paper we will work in the 
ingoing radiation gauge. There are several advantages of choosing this gauge. First, it is straightforward to generalize
the calculations to the Kerr background case. Second, it is expected that the coupling between the  gravitational perturbations
and the scalar field is strong around the event horizon so it is more convenient to use the ingoing radiation gauge. Third,  
the trace of the metric perturbation vanishes in this gauge, $h^{\alpha}_{~\alpha} = 0$. Consequently Eq. (\ref{pHamiltonian}) is simplified: 
\begin{eqnarray}\label{pHamiltonian-2}
H_1 &=& \frac{1}{2} \int_{2M}^{\infty} r^2 \mathrm d r \int_{4\pi} \mathrm d \Omega \big[  -h^{tt} (\partial_t \Phi)^2  + h^{rr} (\partial_r \Phi)^2 \nonumber\\
&& + 2h^{rA}(\partial_r \Phi) (\partial_A \Phi)+ h^{AB}(\partial_A \Phi) (\partial_B \Phi)\big]. 
\end{eqnarray}
In what follows we will consider  the effects of both the odd-parity and even-parity QNMs with frequency $\omega_0 = \omega_R - i \omega_I$ and angular momentum $l_0, m_0$. 

\subsection{Effects of odd-parity QNMs}

\subsubsection{Odd-parity interaction Hamiltonian}\label{OddHamiltonian}

For simplicity, we only consider the coupling between upcoming and upcoming modes, and omit the superscript ``up" without introducing 
any confusion. Couplings between upcoming and ingoing modes, ingoing and ingoing modes are also possible, which we leave for future work.
Since for  odd-parity perturbations, $h^{(o)}_{tt} = h^{(o)}_{rr} = 0$,   the relevant terms in Eq. (\ref{pHamiltonian-2}) are $(\partial_r \Phi) (\partial_A \Phi)$
and $(\partial_A \Phi) (\partial_B \Phi)$.
\begin{widetext}
\begin{eqnarray}\label{drAphi}
(\partial_r\Phi) (\partial_{A}\Phi) &=& \sum_{l=0}^{\infty} \sum_{l'=0}^{\infty} \sum_{m=-l}^l \sum_{m'=-l'}^{l'}  \int \mathrm d \omega \int \mathrm d \omega' 
\frac{1}{4\pi \sqrt{\omega \omega'}} \bigg[ 
\hat{a}_{\omega lm} \hat{a}_{\omega' l'm'} e^{-i (\omega + \omega') t} \partial_r \bigg(\frac{R_{\omega l}}{r} \bigg) \bigg(\frac{R_{\omega' l'}}{r} \bigg)
Y_{lm} (\partial_A Y_{l'm'}) \nonumber \\
&& + \hat{a}_{\omega lm} \hat{a}^{\dag}_{\omega' l'm'} e^{-i (\omega - \omega') t} 
\partial_r \bigg(\frac{R_{\omega l}}{r} \bigg) \bigg(\frac{R^*_{\omega' l'}}{r} \bigg)Y_{lm} (\partial_A Y_{l'm'}^{*}) + \text{h.c.} \bigg].
\end{eqnarray}
\begin{eqnarray}\label{dAAphi}
(\partial_A\Phi) (\partial_{B}\Phi) &=& \sum_{l=0}^{\infty} \sum_{l'=0}^{\infty} \sum_{m=-l}^l \sum_{m'=-l'}^{l'}  \int \mathrm d \omega \int \mathrm d \omega' 
\frac{1}{4\pi \sqrt{\omega \omega'}} \bigg[ 
\hat{a}_{\omega lm} \hat{a}_{\omega' l'm'} e^{-i (\omega + \omega') t} \bigg(\frac{R_{\omega l}}{r} \bigg) \bigg(\frac{R_{\omega' l'}}{r} \bigg)
(\partial_A Y_{lm}) (\partial_B Y_{l'm'}) \nonumber \\
&& + \hat{a}_{\omega lm} \hat{a}^{\dag}_{\omega' l'm'} e^{-i (\omega - \omega') t} 
 \bigg(\frac{R_{\omega l}}{r} \bigg) \bigg(\frac{R^*_{\omega' l'}}{r} \bigg)(\partial_A Y_{lm}) (\partial_B Y_{l'm'}^{*}) + \text{h.c.} \bigg].
\end{eqnarray}
\end{widetext}
From Eq. (\ref{OddPerturbation}) we see that $h^{(o)}_{rA}$ and $h^{(o)}_{AB}$ contain terms that are proportional to 
$e^{-i\omega_0 t} = e^{-\omega_I t} e^{-i\omega_R t}$ and $(e^{-i\omega_0 t})^* = e^{-\omega_I t} e^{i\omega_R t}$.
When multiplying with $(\partial_r \Phi) (\partial_A \Phi)$ and $(\partial_A \Phi) (\partial_B \Phi)$ we get terms containing
factors 
\begin{eqnarray}
&&e^{\pm i (\omega_R - \omega - \omega') t}, ~~ e^{\pm i (\omega_R + \omega + \omega') t}, \nonumber\\
&&e^{\pm i (\omega_R + \omega - \omega') t}, ~~ e^{\pm i (\omega_R - \omega + \omega') t}. \nonumber
\end{eqnarray}
In the rotating-wave approximation, terms with the lowest frequency oscillations $e^{\pm i (\omega_R - \omega - \omega') t}$ dominate over more highly  oscillatory terms. This approximation ensures that the energy is approximately conserved, $\omega_R \approx \omega + \omega'$.
Substituting Eqs. (\ref{OddPerturbation}),  (\ref{drAphi}) and  (\ref{dAAphi})  into Eq. (\ref{pHamiltonian-2}) we have in this approximation
\begin{widetext}
\begin{eqnarray}\label{pHamiltonian-odd}
H^{(o)}_1 \approx \frac{1}{16\pi} e^{-\omega_I t} \sum_{l=0}^{\infty} \sum_{l'=0}^{\infty} \sum_{m=-l}^l \sum_{m'=-l'}^{l'}  \int \mathrm d \omega \int \mathrm d \omega' 
\frac{1}{\sqrt{\omega \omega'}} \bigg[ 
e^{- i (\omega_R - \omega - \omega') t} \hat{a}_{\omega lm}^{\dag} \hat{a}_{\omega' l'm'}^{\dag} 
(\mathcal{I}_{r1} \mathcal{I}_{a1} + \mathcal{I}_{r2} \mathcal{I}_{a2}) + \text{h.c.}
\bigg],
\end{eqnarray}
where $\mathcal{I}_{r1} $ and $\mathcal{I}_{r2}$ are the radial integrals,
\begin{eqnarray}
\mathcal{I}_{r1} &=& - \frac{\sqrt{D_0}}{\sqrt{2l_0(l_0+1)}} \int_{2M}^{\infty} \mathrm d r \frac{1}{f} \label{RI1}
\bigg[\frac{d}{d r_*} {}_{-2}R_{\omega_0 l_0}- \bigg(i\omega_0 + \frac{2 f}{r}  \bigg) {}_{-2}R_{\omega_0 l_0}\bigg]
\partial_r \bigg(\frac{R^*_{\omega l}}{r} \bigg) \bigg(\frac{R^*_{\omega' l'}}{r} \bigg), \\
\mathcal{I}_{r2}&=&  \int_{2M}^{\infty} \mathrm d r \frac{1}{r^2 f^2} 
\bigg[(i \omega_0 r^2 - M)\frac{d}{d r_*}{}_{-2}R_{\omega_0 l_0} - \bigg(\frac{1}{2} \mu_0^2 f - i\omega_0 (-3r+7M) - \omega_0^2 r^2  \bigg){}_{-2}R_{\omega_0 l_0}\bigg]
\bigg(\frac{R^*_{\omega l}}{r} \bigg) \bigg(\frac{R^*_{\omega' l'}}{r} \bigg),
\end{eqnarray}
$\mathcal{I}_{a1} $ and $\mathcal{I}_{a2}$ are the angular integrals,
\begin{eqnarray}
\mathcal{I}_{a1} (l_0, m_0; l, m; l', m')
&=& \int_{4\pi} \mathrm d \Omega \big({}_{-1}Y_{l_0 m_0} m^A + {}_{1}Y_{l_0 m_0} m^{A*} \big) Y^*_{lm} (\partial_A Y^*_{l'm'}), \label{AI1}\\
\mathcal{I}_{a2}  (l_0, m_0; l, m; l', m')
&=& \int_{4\pi} \mathrm d \Omega  \big({}_{-2}Y_{l_0 m_0} m^A  m^B - {}_{2}Y_{l_0 m_0} m^{A*} m^{B*} \big) (\partial_A Y^*_{lm}) (\partial_B Y^*_{l'm'}). \label{AI2}
\end{eqnarray}
\end{widetext}
Since the Hamiltonian  (\ref{pHamiltonian-odd}) is quadratic in creation (and annihilation) operators, it is clear that it describes multimode squeezing.   The quantity
$\mathcal{I}_{r1} \mathcal{I}_{a1} + \mathcal{I}_{r2} \mathcal{I}_{a2} $ plays the role of a phase matching function,
the nonzero value of which would imply that  gravitational perturbations generate quantum particles.


%
%
\subsubsection{Radial and angular integrals}\label{RAintegral}

Since there are no analytic solutions for the radial functions $R_{\omega l}$ and ${}_{-2}R_{\omega_0 l_0}$, it is therefore impossible to find analytic results for
the radial integrals $\mathcal{I}_{r1}$ and $\mathcal{I}_{r2}$. One might expect the radial integrals can be calculated numerically. It turns out that the calculation of
the radial integrals is not trivial because of the peculiar property of the radial function of the QNMs. From the boundary conditions for the QNMs, Eq. (\ref{boundary:QNM}), 
the radial function of the QNMs is proportional to $e^{\omega_I r_*}$ when $r_* \rightarrow +\infty$; and it is proportional to $e^{-\omega_I r_*}$ when $r_* \rightarrow -\infty$.
In both limits the radial function of the QNMs is divergent and the radial integrals are not  well defined. Leaver \cite{Leaver86} proposed a method to overcome this difficulty by exploiting the analyticity of the integrand in $r_*$. A new contour (see Appendix \ref{appendixA} for details) is chosen such that the integral along this contour is finite. 
Sun and Price \cite{SunPrice88} discussed in detail how to construct Leaver's contour by analytic continuation and restored a factor that is missed in \cite{Leaver86}.
Similar techniques were also used by Yang {\it et al} \cite{Yang15} to define the inner product of the radial function of the QNMs. In this paper, we follow the method of 
Leaver (taking into account the missing factor) to regularize the radial integral to obtain a finite result.  

By using Leaver's method, the radial integral can be in principle calculated numerically. In order to obtain an approximately analytic result, we assume that the main 
contribution to the integration is from the region near the event horizon, that is, $r \sim 2M$. This is because the coupling between the QNMs and scalar field near the horizon is 
expected to be stronger. This assumption can be also justified by looking at the asymptotic behaviour of the integrand along the contour at infinity, which is exponentially 
suppressed (see Appendix \ref{appendixA} for details). To further simplify the result, we assume that 
the imaginary part of the QNM frequencies are small. This is rather a crude 
approximation because the imaginary part of the QNM frequencies of a Schwarzschild black hole are not so small. However this approximation is adequate for the purpose of this 
paper. 

Taking into account all the above approximations, we find 
\begin{equation}\label{Rintegral:1}
\mathcal{I}_{r1} \approx - \frac{\sqrt{D_0}}{\sqrt{2l_0(l_0+1)}} \frac{16 \pi i M \Omega \Omega_0 }{(i \Delta)^2} 
\frac{A^*_{\omega l} A^*_{\omega' l'}  C_{\omega_0 l_0}}{1 - e^{2\pi \Omega_0}},
\end{equation}
and 
\begin{equation}\label{Rintegral:2}
\mathcal{I}_{r2} \approx \frac{16 \pi i M \Omega_0^2 }{(i \Delta)^2} \frac{A^*_{\omega l} A^*_{\omega' l'}  C_{\omega_0 l_0}}{1 - e^{2\pi \Omega_0}},
\end{equation}
where  $\Omega = 2M \omega$, $\Omega_0 = 2M \omega_0$, $\Omega' = 2M \omega'$ and $\Delta = \Omega_0 - \Omega - \Omega'$.
Eqs. (\ref{Rintegral:1}) and (\ref{Rintegral:2}) are valid when $\Delta \sim 0$. 

Fortunately, the angular integrals $\mathcal{I}_{a1}$ and $\mathcal{I}_{a2}$ can be calculated analytically. In Appendix \ref{appendix:SWS}, 
we derive the explicit expressions for the spin-weighted spherical harmonics with spin weight $\pm 1$ and $\pm 2$ in terms of the Legendre functions, 
as shown by Eqs. (\ref{SWH1}) and (\ref{SWH2}). In Appendix \ref{appendix:angular12},
we substitute Eqs. (\ref{SWH1}) and (\ref{SWH2}) into Eqs. (\ref{AI1}) and (\ref{AI2}) to obtain the angular integrals $\mathcal{I}_{a1}$ and $\mathcal{I}_{a2}$. 

For simplicity we consider a special case where the angular momentum of the QNMs along the $z$ direction is zero, that is, $m_0 = 0$. 
This simplifies the calculation a lot and is sufficient to demonstrate quantum particle generation by the QNMs. The resulting angular integrals 
$\mathcal{I}_{a1}$ and $\mathcal{I}_{a2}$ are given by Eqs. (\ref{AngularIntegral:1}) and (\ref{AngularIntegral:2}).
Note that the integration over $\phi$ gives rise to a $\delta$-function $\delta_{m', -m}$, which implies that the producing pair of 
particles have opposite angular momentum along the $z$ direction. This is not surprising given that $m_0=0$ and is 
an indication of the angular momentum conservation in the particle production process.

As an example, we calculate the 
angular integrals $\mathcal{I}_{a1}$ and $\mathcal{I}_{a2}$ for a QNM with angular momentum $(l_0, m_0)=(2, 0)$ and a pair of scalar particles, 
the first of which with angular momentum $(l, m)=(1, 1)$ and the other $(l', -1)$. We find that the particle $(1, 1)$ only couples with those particles with  $l' =2$, namely, the only
nonzero $\mathcal{I}_{a1}$ and $\mathcal{I}_{a2}$ are
\begin{eqnarray}
\mathcal{I}_{a1} (2, 0; 1, 1; 2, -1) &=& -\frac{1}{2} \sqrt{\frac{3}{\pi}}, \nonumber\\
\mathcal{I}_{a2} (2, 0; 1, 1; 2, -1) &=& \sqrt{\frac{3}{2\pi}}.
\end{eqnarray}

\subsubsection{QNM as multimode squeezer}

In Sec. \ref{RAintegral}, we have analytically calculated the angular integrals, and derived approximately analytic expressions for the radial integrals
for odd-parity QNMs. We thus can obtain the interaction Hamiltonian $H_1^{\text{(o)}}$, which dominates the evolution of the scalar field. 
In the following we will estimate the strength of the coupling between the QNMs and the scalar field. 


The time evolution operator is 
\begin{equation}
\hat{U}^{(o)} = \hat{T} \exp \bigg\{-i \int_0^{\infty} \mathrm d t ~ H_1^{\text{(o)}} (t) \bigg\},
\end{equation}
where $\hat{T}$ is the time ordering operator. In the low squeezing regime \cite{Wasilewski06}, the time ordering is not important so that we can approximate the time
evolution operator as 
\begin{equation}
\hat{U}^{(o)} \approx \exp \bigg\{-i \int_0^{\infty} \mathrm d t ~ H_1^{\text{(o)}} (t) \bigg\},
\end{equation}
and the  integration can be directly carried out. 
Using  
\begin{equation*}
\int_0^{\infty} \mathrm d t~ e^{-\omega_I t} e^{-i (\omega_R-\omega-\omega') t} = \frac{1}{i \big[(\omega_R-\omega-\omega') - i \omega_I \big]},
\end{equation*}
we have 
\begin{eqnarray}\label{SqueezingOperator}
\hat{U}^{(o)} &=& \bigotimes_{lm} \bigotimes_{l'} \hat{U}^{(o)}_{lml'}, \nonumber \\
\hat{U}^{(o)}_{lml'} &\approx& \exp \bigg\{-i \int \mathrm d \omega \int \mathrm d \omega' 
\bigg[\mathcal{F}^{(o)}_{lml'} (\omega, \omega') \hat{a}^{\dag}_{\omega lm} \hat{a}^{\dag}_{\omega' l', -m} \nonumber\\
&&+ \mathcal{F}^{(o)*}_{lml'} (\omega, \omega') \hat{a}_{\omega lm} \hat{a}_{\omega' l', -m} \bigg] \bigg\} 
\end{eqnarray}
where 
\begin{eqnarray}\label{JointDistribution}
\mathcal{F}^{(o)}_{lml'}(\omega, \omega') = \frac{4 i M^3 A^*_{\omega l} A^*_{\omega' l'} C_{\omega_0 l_0} }{\sqrt{\Omega \Omega'} (1 - e^{2\pi \Omega_0})} 
\frac{1}{(i \Delta)^3} \nonumber\\
\times \bigg[\Omega_0^2\mathcal{I}_{a2} - \frac{\sqrt{D_0}}{\sqrt{2l_0(l_0+1)}} \Omega \Omega_0 \mathcal{I}_{a1} \bigg]. 
\end{eqnarray}
It is evident that Eq. (\ref{SqueezingOperator}) represents a multimode squeezing operator and 
$\mathcal{F}^{(o)}_{lml'}(\omega, \omega')$ is known as the joint frequency distribution. Eq. (\ref{JointDistribution}) shows that the joint frequency distribution 
$\mathcal{F}^{(o)}_{lml'}(\omega, \omega')$ is not zero, indicating that there will be scalar particle creation. If the initial state of the scalar field is a vacuum state (the  Boulware vacuum), the QNMs  squeeze the vacuum and produce a squeezed vacuum state; if the initial state is a thermal state, e.g., Hawking thermal radiation, it is amplified due to the 
QNM   squeezing. In general the QNMs squeeze any state of the scalar field and amplify it,  producing scalar  particles. The particle production energy  comes from the QNMs. This is our main result.
 
\begin{figure}[ht!]

\centering
\begin{tikzpicture}[scale=0.66667]
  \node[anchor=south west,inner sep=0] at (0,0) {\includegraphics[width=8 cm]{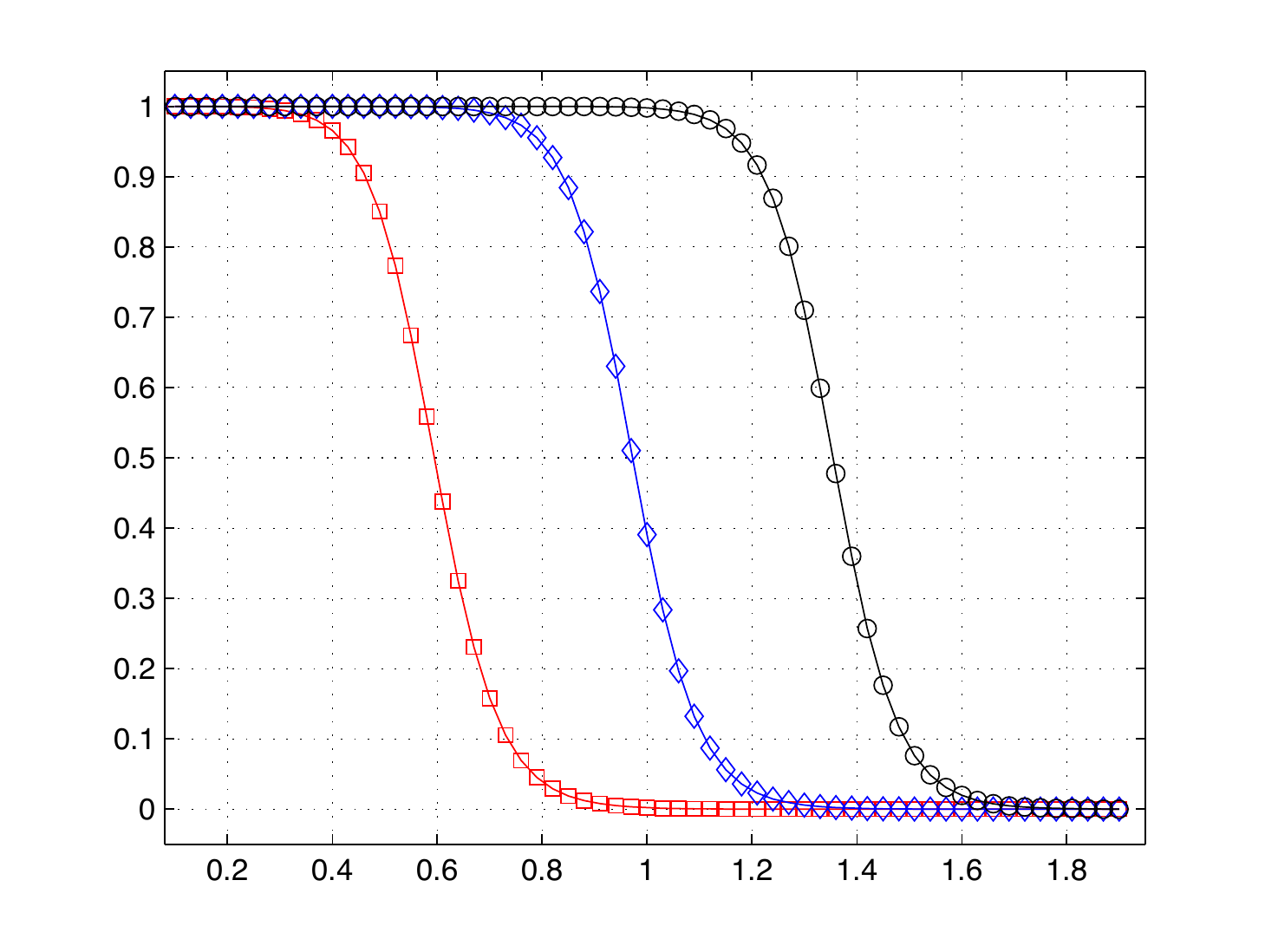}};
     
     \draw  (6, 0.2) node {$2M\omega$};
     \draw  (0.5, 4.5) node [rotate=90] {\large $|A_{\omega l}|^2$};
     
     
     \draw [color=red] (4.8, 4.8) node {$l=1$};
     \draw [color=blue] (6.7, 4.8) node {$l=2$};
     \draw (8.6, 4.8) node {$l=3$}; 
     
\end{tikzpicture}
\caption{\footnotesize (colour online) Reflection coefficient for the scalar field modes.} 
\label{fig:Reflection}
\end{figure}
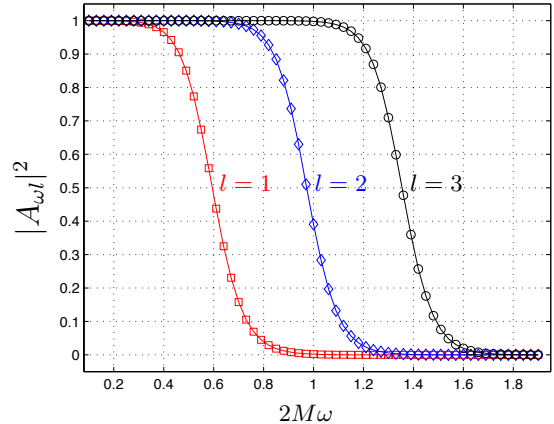

In Eq. (\ref{JointDistribution}), $A_{\omega l}$ is the reflection amplitude (see Eq. (\ref{upasymptotic})) of the upcoming scalar field mode determined by the 
effective potential  (\ref{EffPot:Scalar}). 
For a given $l$, the effective potential peaks around $r_* = 2M$ with its maximum
depending on $l$ (higher for larger $l$). 

Unfortunately, there is no analytic expression for the reflection amplitude.  However we can infer the qualitative behaviour of $A_{\omega l}$. When the frequency of the field mode is lower than the maximum of the effective potential, most of the field mode is reflected, whereas
the field mode almost penetrates through the potential  if the frequency is higher than the potential maximum. Fig. \ref{fig:Reflection} shows the numerical results of the reflection coefficients, 
$|A_{\omega l}|^2$.
%
%
%
%
%
%

\begin{figure}[ht!]

\begin{tikzpicture}[scale=0.9]
  \node[anchor=south west,inner sep=0] at (0,0) {\includegraphics[width=9 cm]{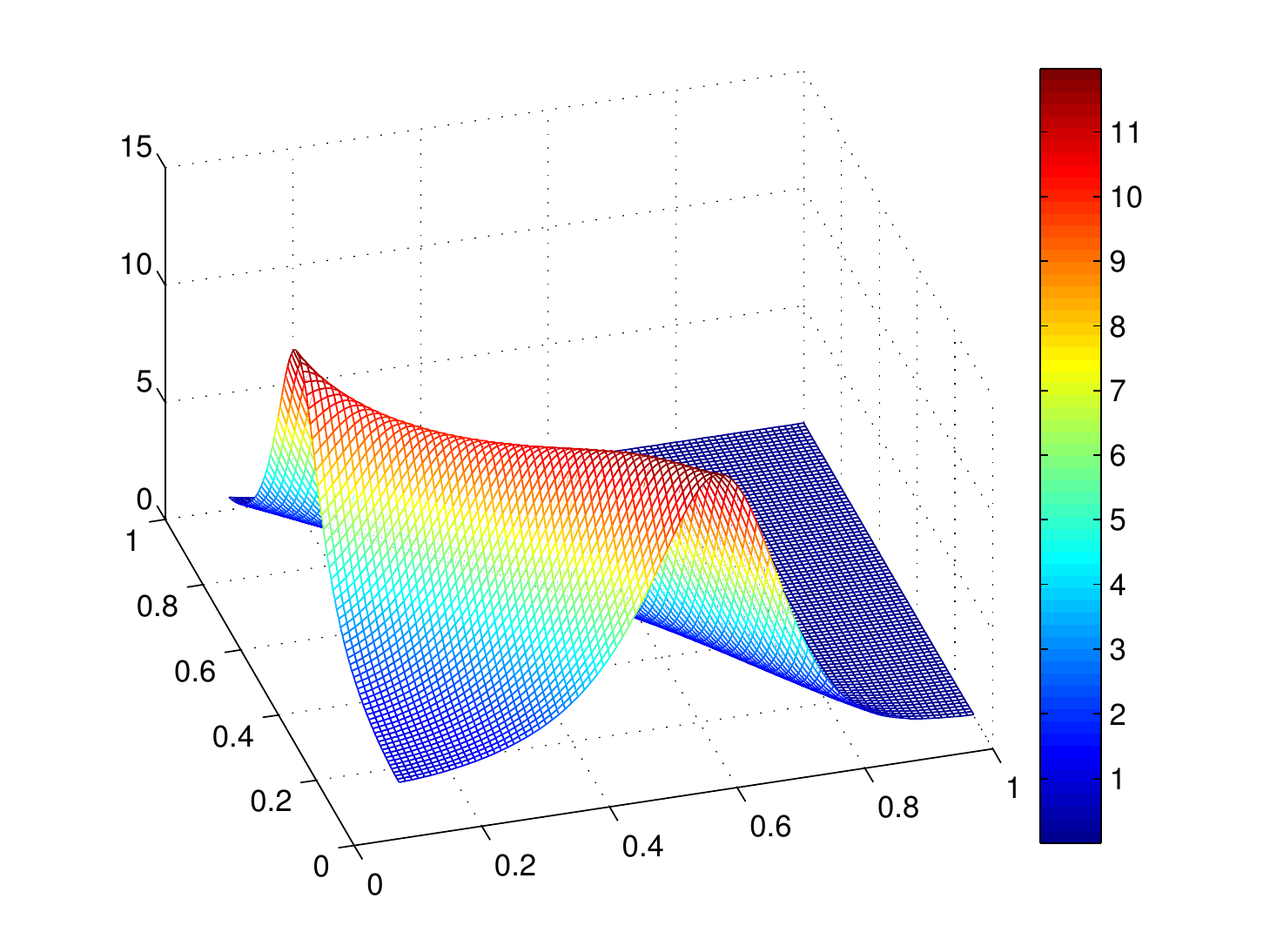}};
     
     \draw (1.2, 1.8) node [rotate=-62] {$2M\omega'$};
     \draw  (5.5, 0.5) node  [rotate= 6] {$2M\omega$};
     \draw  (0.2, 4.8) node [rotate=90] {$|\mathcal{F}^{(o)}_{lml'}(\omega, \omega')|/M^3|C_{\omega_0 l_0}|$};

\end{tikzpicture}
\caption{\footnotesize (colour online) Modulus of the joint frequency distribution. QNM: $\Omega_R = 0.7474, \Omega_I = 0.178$, $(l_0, m_0) = (2, 0)$. 
Scalar particle one: $(l, m) = (1, 1)$; scalar particle two: $(l', m') = (2, -1)$}
\label{fig:JFD-1}
\end{figure}
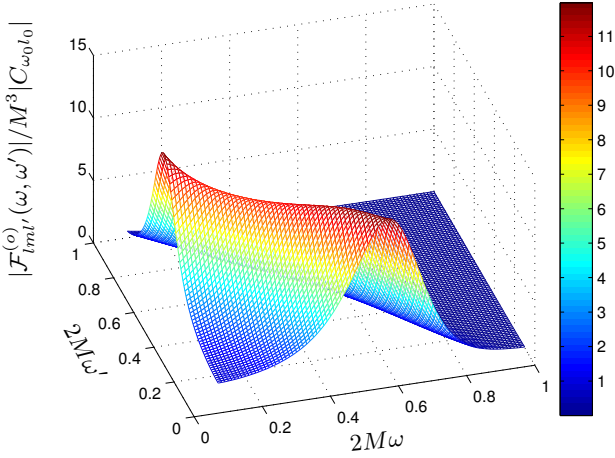

After knowing the behaviour of the reflection amplitude, we can study the behaviour of the joint frequency distribution. The factor $\Delta$ is defined as
\begin{eqnarray*}
\Delta \equiv \Omega_0 - \Omega - \Omega' = (\Omega_R-\Omega-\Omega') - i \Omega_I,
\end{eqnarray*}
where $\Omega_R = 2M \omega_R$, $\Omega_I = 2M \omega_I$. Therefore we have 
\begin{eqnarray}\label{FrequencyDistribution}
\frac{1}{|\Delta|^3} = \frac{1}{\big[(\Omega_R-\Omega-\Omega')^2 + \Omega_I^2\big]^{3/2}}. 
\end{eqnarray}
This is a distribution with respect to $\Omega$ and $\Omega'$ that peaks along the line $\Omega+\Omega' = \Omega_R$, the maximum of which 
is $1/\Omega_I^3$. Here $\Omega_R-\Omega-\Omega'$ can be considered as the frequency detuning, and $\Omega_I$ can be considered as the 
decay rate which also characterizes the width of the distribution Eq. (\ref{FrequencyDistribution}). If $\Omega_I$ is small, the distribution Eq. (\ref{FrequencyDistribution})
is nonzero only for $\Omega+\Omega' \approx \Omega_R$. This is an indication of energy conservation: the sum of the frequencies of the pair of 
scalar particles is equal to the real part of the QNM frequency. Figs. \ref{fig:JFD-1} 
shows an example of the absolute value of the joint frequency distribution, 
$|\mathcal{F}^{(o)}_{lml'}(\omega, \omega')|$. We can see that basically $|\mathcal{F}^{(o)}_{lml'}(\omega, \omega')|$ follows the energy-conservation line 
$\Omega+\Omega' = \Omega_R$. The high frequency part is suppressed by the reflection amplitude $A_{\omega l}$; while in the low frequency regime, $|A_{\omega l}|$
is almost one and the factor $1/\sqrt{\Omega \Omega'}$ dominates. The latter is annoying because that means the joint frequency distribution is divergent 
at $\Omega=0$ or $\Omega'=0$.  Here we assume that there exists a low frequency cutoff so that the joint frequency distribution is finite. 

%
%
%

We can compute a crude estimate of the maximum of the joint frequency distribution. Assume that in Eq. (\ref{JointDistribution}), $\Omega \sim \Omega' \sim \Omega_R/2$, 
$|A_{\omega l}| \sim |A_{\omega' l'}| \sim 1/\sqrt{2}$ and the contribution from the angular integral part is at the order of unity, we find
\begin{eqnarray}\label{MaxSqueezing}
|\mathcal{F}^{(o)}_{lml'}|_{\text{max}} \sim 4 M^3 \bigg( \frac{\Omega_R}{\Omega_I} \bigg) \bigg( \frac{1}{\Omega^2_I} \bigg) 
\frac{|C_{\omega_0 l_0}|}{\big|1 - e^{2\pi (\Omega_R - i \Omega_I)} \big|}. 
\end{eqnarray}
This is an approximate relation between the squeezing amplitude (or the coupling strength) and various parameters of the black hole and the QNM.
Here $\Omega_R$ is the resonance frequency of the QNM and $\Omega_I$ characterizes the decay rate. If we make an analogy with an optical cavity \cite{Bachor04}, 
$\Omega_R/\Omega_I$ can be considered as the quality factor of the QNM. We see that the squeezing amplitude is proportional to the cube of the black hole mass,
the amplitude of the QNM at the event horizon, the quality factor of the QNM, and is inversely proportional to the square of the decay rate. 
This means the coupling strength is stronger for longer lasting QNMs, larger amplitude QNMs and 
bigger black holes. However for Schwarzschild black holes the QNMs decay very fast and the least damped QNM is the fundamental QNM for which the overtone number 
is $n=0$ \cite{Berti09}. For example, for the fundamental QNM of $(l_0, m_0) = (2, 0)$, $\Omega_R = 0.7474$ and $\Omega_I = 0.178$ \cite{Berti09}. Substituting these into
Eq. (\ref{MaxSqueezing}), we find
\begin{eqnarray}
|\mathcal{F}^{(o)}_{lml'}|_{\text{max}} \sim 5 M^3 |C_{\omega_0 l_0}|. 
\end{eqnarray}

\subsection{Effects of even-parity QNMs}

The procedure to evaluate the coupling between the scalar field and the even-parity QNMs is similar.
For the even-parity perturbations, $h^{(e)}_{tt} \ne 0$, $h^{(e)}_{rr} \ne 0$, so we also need $(\partial_t \Phi) (\partial_t \Phi)$  and $(\partial_r \Phi) (\partial_r \Phi)$
in Eq. (\ref{pHamiltonian-2}), which is 
\begin{widetext}
\begin{eqnarray}
(\partial_t \Phi) (\partial_{t}\Phi) &=& \sum_{l=0}^{\infty} \sum_{l'=0}^{\infty} \sum_{m=-l}^l \sum_{m'=-l'}^{l'}  \int \mathrm d \omega \int \mathrm d \omega' \frac{1}{2\pi \sqrt{\omega \omega'}} \bigg[ 
- \omega \omega' \hat{a}_{\omega lm} \hat{a}_{\omega' l'm'} e^{-i (\omega + \omega') t} \bigg(\frac{R_{\omega l}}{r} \bigg) \bigg(\frac{R_{\omega' l'}}{r} \bigg)
Y_{lm} Y_{l'm'} \nonumber \\
&& + \omega \omega' \hat{a}_{\omega lm} \hat{a}^{\dag}_{\omega' l'm'} e^{-i (\omega - \omega') t} 
\bigg(\frac{R_{\omega l}}{r} \bigg) \bigg(\frac{R^*_{\omega' l'}}{r} \bigg)Y_{lm} Y_{l'm'}^{*} + \text{h.c.} \bigg].
\end{eqnarray}
\begin{eqnarray}\label{drdr}
(\partial_r\Phi) (\partial_{r}\Phi) &=& \sum_{l=0}^{\infty} \sum_{l'=0}^{\infty} \sum_{m=-l}^l \sum_{m'=-l'}^{l'}  \int \mathrm d \omega \int \mathrm d \omega' 
\frac{1}{4\pi \sqrt{\omega \omega'}} \bigg[ 
\hat{a}_{\omega lm} \hat{a}_{\omega' l'm'} e^{-i (\omega + \omega') t} \partial_r \bigg(\frac{R_{\omega l}}{r} \bigg) \partial_r \bigg(\frac{R_{\omega' l'}}{r} \bigg)
Y_{lm} Y_{l'm'} \nonumber \\
&& + \hat{a}_{\omega lm} \hat{a}^{\dag}_{\omega' l'm'} e^{-i (\omega - \omega') t} 
\partial_r \bigg(\frac{R_{\omega l}}{r} \bigg) \partial_r \bigg(\frac{R^*_{\omega' l'}}{r} \bigg)Y_{lm} Y_{l'm'}^{*} + \text{h.c.} \bigg].
\end{eqnarray}
Substituting Eqs. (\ref{EvenPerturbation}) and (\ref{ScalarField}) into  (\ref{pHamiltonian-2}), and taking into account the rotating wave
approximation, we have 
\begin{eqnarray}\label{pHamiltonian-even}
H^{(e)}_1 &\approx& \frac{1}{8\pi} e^{-\omega_I t} \sum_{l=0}^{\infty} \sum_{l'=0}^{\infty} \sum_{m=-l}^l \sum_{m'=-l'}^{l'}  \int \mathrm d \omega \int \mathrm d \omega' \frac{1}{\sqrt{\omega \omega'}} 
\bigg\{
e^{- i (\omega_R - \omega - \omega') t} \hat{a}_{\omega lm}^{\dag} \hat{a}_{\omega' l'm'}^{\dag} 
\big[(\mathcal{I}_{r3} + \mathcal{I}_{r4}) \mathcal{I}_{a3}+\mathcal{I}_{r1} \mathcal{I}_{a4} + \mathcal{I}_{r2} \mathcal{I}_{a5}\big] \nonumber\\
&& + \text{h.c.} \bigg\},
\end{eqnarray}
where the radial integrals $\mathcal{I}_{r1} $ and $\mathcal{I}_{r2} $ are defined in Sec. \ref{OddHamiltonian}, $\mathcal{I}_{r3}$ and $\mathcal{I}_{r4}$ are defined as
\begin{eqnarray}
\mathcal{I}_{r3} &=& - \omega \omega' \sqrt{D_0} \int_{2M}^{\infty} \mathrm d r \frac{1}{f^2} \label{R31}
{}_{-2}R_{\omega_0 l_0} \bigg(\frac{R^*_{\omega l}}{r} \bigg) \bigg(\frac{R^*_{\omega' l'}}{r} \bigg), \\
\mathcal{I}_{r4} &=& - \sqrt{D_0} \int_{2M}^{\infty} \mathrm d r \label{R31}
{}_{-2}R_{\omega_0 l_0} \partial_r \bigg(\frac{R^*_{\omega l}}{r} \bigg) \partial_r \bigg(\frac{R^*_{\omega' l'}}{r} \bigg).
\end{eqnarray}
The angular integrals $\mathcal{I}_{a3}$, $\mathcal{I}_{a4}$, $\mathcal{I}_{a5}$ are
\begin{eqnarray}
\mathcal{I}_{a3} (l_0, m_0; l, m; l', m') &=& \int_{4\pi} \mathrm d \Omega ~Y_{l_0 m_0} Y^*_{lm} Y^*_{l'm'}, \\ 
\mathcal{I}_{a4} (l_0, m_0; l, m; l', m') &=& \int_{4\pi} \mathrm d \Omega ~\big({}_{-1}Y_{l_0 m_0} m^A - {}_{1}Y_{l_0 m_0} m^{A*} \big) Y^*_{lm} (\partial_A Y^*_{l'm'}), \\
\mathcal{I}_{a5} (l_0, m_0; l, m; l', m') &=& \int_{4\pi} \mathrm d \Omega  ~\big({}_{-2}Y_{l_0 m_0} m^A  m^B + {}_{2}Y_{l_0 m_0} m^{A*} m^{B*} \big) (\partial_A Y^*_{lm}) (\partial_B Y^*_{l'm'}). \label{A32}
\end{eqnarray}
\end{widetext}
The radial integral $\mathcal{I}_{r3}$ and $\mathcal{I}_{r4}$ can be calculated  similarly  to that for $\mathcal{I}_{r1}$; details are in
Appendix \ref{appendixA}. In the limit   $i\Delta \sim 0$, we find
\begin{equation}
\mathcal{I}_{r3} \approx -\mathcal{I}_{r4}\approx - \frac{8 \pi i \sqrt{D_0} M \Omega \Omega' }{(i \Delta)^2} 
\frac{A^*_{\omega l} A^*_{\omega' l'} C_{\omega_0 l_0}}{1 - e^{2\pi \Omega_0}}. 
\end{equation}
This implies $\mathcal{I}_{r3} + \mathcal{I}_{r4} \approx 0$. 

It turns out that $\mathcal{I}_{a3}$ can be easily obtained and expressed in terms of the 3-$j$ symbols,
\begin{eqnarray}
\mathcal{I}_{a3}  &=& (-1)^{m_0}\sqrt{\frac{(2l_0 + 1)(2l + 1)(2l' + 1)}{4\pi}}
\begin{pmatrix}
l & l' & l_0 \\ 
0 & 0  & 0 
\end{pmatrix} \nonumber \\
&&\times
\begin{pmatrix}
l & l' & l_0 \\ 
-m & -m'  & m_0 
\end{pmatrix},
\end{eqnarray}
which is zero when $m_0 \ne m + m'$. Taking into account the properties of the 3-$j$ symbols, we find that $\mathcal{I}_{a3}$
vanishes when $l_0 + l + l'$ is an odd integer. 

If we consider the special case where the angular momentum of the QNMs along the $z$ direction is zero, that is, $m_0=0$,
the calculation can be significantly simplified. The resulting angular integrals 
$\mathcal{I}_{a4}$ and $\mathcal{I}_{a5}$ are given by Eqs. (\ref{AngularIntegral:4}) and (\ref{AngularIntegral:5}).
As an example, we calculate the 
angular integrals $\mathcal{I}_{a4}$ and $\mathcal{I}_{a5}$ for a QNM with angular momentum $(l_0, m_0)=(2, 0)$ and a pair of scalar particles, 
the first of which with angular momentum $(l, m)=(1, 1)$ and the other $(l', -1)$. We find that the only nonzero $\mathcal{I}_{a4}$ and $\mathcal{I}_{a5}$ are for 
$l' = 1$ and $l' = 3$, as shown in Table \ref{tab:AI45}.
\begin{table}
  \centering
  \scalebox{1.2}{
  \linespread{2}\selectfont
  \begin{tabular}{|p{2.8 cm}|p{1.5 cm}|p{1.5 cm}|}
    \hline
      & $l'=1$ & $l'=3$  \\
    \hline
    $\mathcal{I}_{a4}(2,0; 1, 1; l', -1)$ & $\frac{1}{2}\sqrt{\frac{3}{5\pi}}$ & $-4 \sqrt{\frac{6}{35\pi}}$  \\
    \hline
    $\mathcal{I}_{a5}(2,0; 1, 1; l', -1)$ & $\sqrt{\frac{3}{10\pi}}$ & $2\sqrt{\frac{3}{35\pi}}$ \\
    \hline
  \end{tabular}
  }
  \caption{Nonzero $\mathcal{I}_{a4}$ and $\mathcal{I}_{a5}$ for a QNM $(l_0, m_0) = (2, 0)$ and a pair of scalar particles: $(l, m) = (1, 1)$ and $(l', -1)$.  }
  \label{tab:AI45}
\end{table}

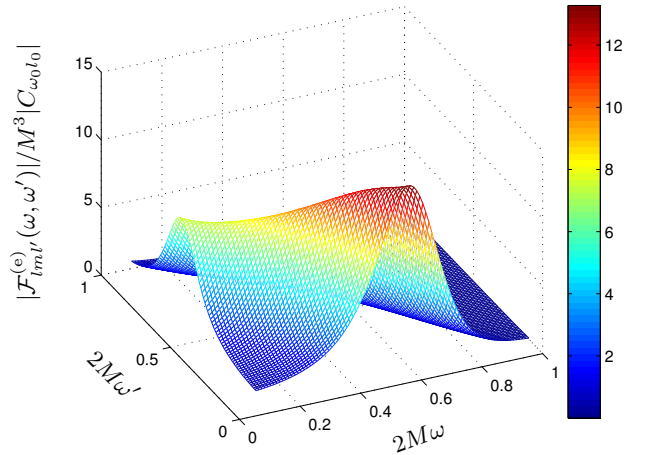
\begin{figure}[ht!]

\centering
\begin{tikzpicture}[scale=0.9]
  \node[anchor=south west,inner sep=0] at (0,0) {\includegraphics[width=9 cm]{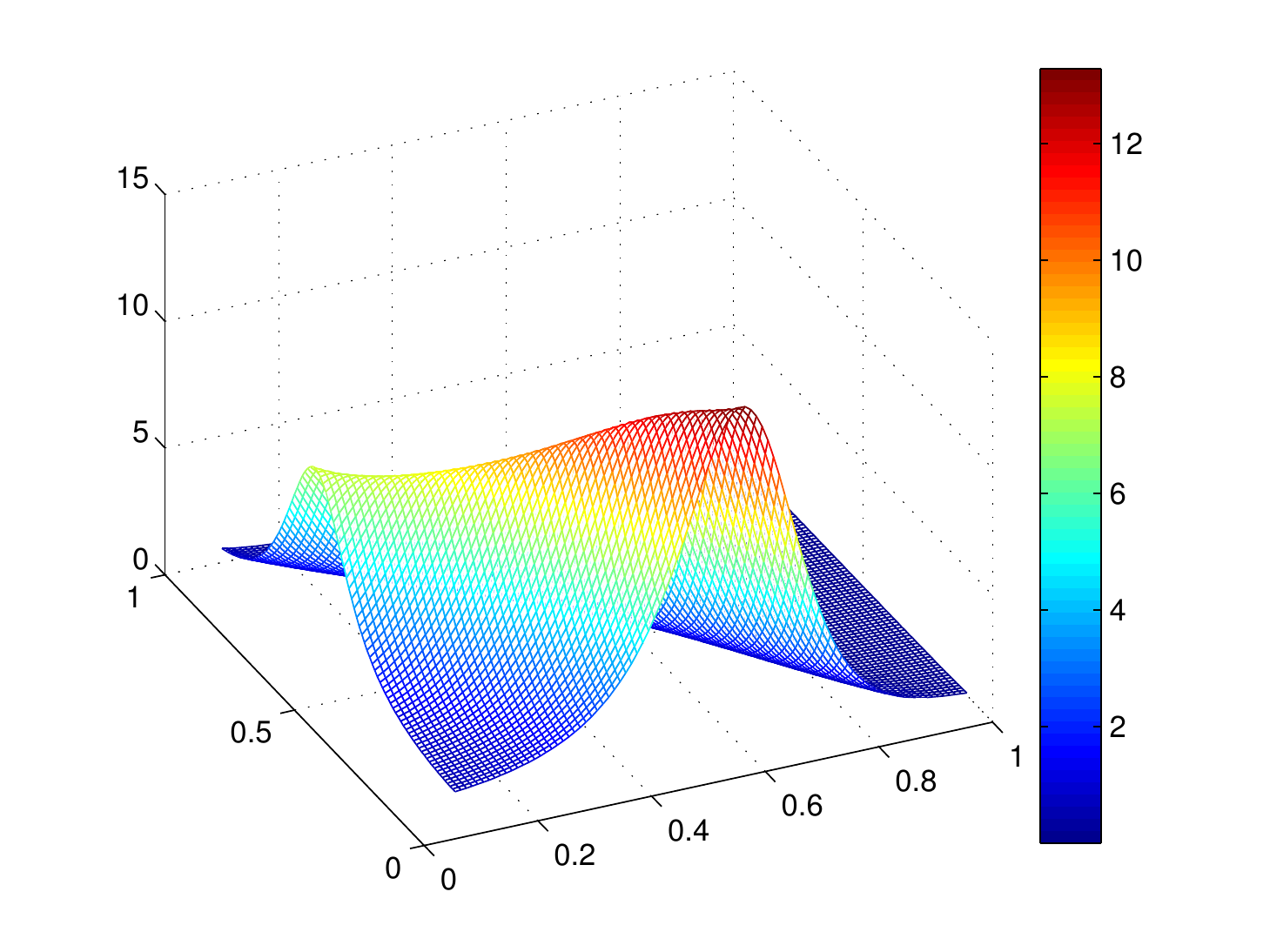}};
     
     \draw (1.5, 1.5) node [rotate=-45] {$2M\omega'$};
     \draw  (6., 0.6) node  [rotate= 20] {$2M\omega$};
     \draw  (0.2, 4.5) node [rotate=90] {$|\mathcal{F}^{\text{(e)}}_{lml'}(\omega, \omega')|/M^3|C_{\omega_0 l_0}|$};

\end{tikzpicture}
\caption{\footnotesize (colour online) Modulus of the joint frequency distribution for even-parity QNMs. QNM: $\Omega_R = 0.7474, \Omega_I = 0.178$, $(l_0, m_0) = (2, 0)$. 
Scalar particle one: $(l, m) = (1, 1)$; scalar particle two: $(l', m') = (3, -1)$.}
\label{fig:JFD-even-1}
\end{figure}

The time evolution operator is 
\begin{equation}
\hat{U}^{\text{(e)}} = \hat{T} \exp \bigg\{-i \int_0^{\infty} \mathrm d t ~ H_1^{\text{(e)}} (t) \bigg\},
\end{equation}
where $\hat{T}$ is the time ordering operator; as before, in the low downconversion regime \cite{Wasilewski06} time ordering is not important  and  we find
\begin{eqnarray}\label{SqueezingOperator:Even}
\hat{U}^{\text{(e)}} &=& \bigotimes_{lm} \bigotimes_{l'} \hat{U}^{\text{(e)}}_{lml'}, \nonumber \\
\hat{U}^{\text{(e)}}_{lml'} &\approx& \exp \bigg\{-i \int \mathrm d \omega \int \mathrm d \omega' 
\bigg[\mathcal{F}^{\text{(e)}}_{lml'} (\omega, \omega') \hat{a}^{\dag}_{\omega lm} \hat{a}^{\dag}_{\omega' l', -m} \nonumber\\
&&+ \mathcal{F}^{\text{(e)}*}_{lml'} (\omega, \omega') \hat{a}_{\omega lm} \hat{a}_{\omega' l', -m} \bigg] \bigg\} 
\end{eqnarray}
where the joint frequency distribution is
\begin{eqnarray}\label{JointDistribution:Even}
\mathcal{F}^{(e)}_{lml'}(\omega, \omega') = \frac{4 i M^3 A^*_{\omega l} A^*_{\omega' l'} C_{\omega_0 l_0} }{\sqrt{\Omega \Omega'} (1 - e^{2\pi \Omega_0})} 
\frac{1}{(i \Delta)^3} \nonumber\\
\times \bigg[\Omega_0^2\mathcal{I}_{a5} - \frac{\sqrt{D_0}}{\sqrt{2l_0(l_0+1)}} \Omega \Omega_0 \mathcal{I}_{a4} \bigg]. 
\end{eqnarray}
It is evident that the joint frequency distribution $\mathcal{F}^{\text{(e)}}_{lml'}(\omega, \omega')$ for the coupling between the scalar field and the even-parity 
QNM is nonzero, showing that there is scalar particle creation. Figs. \ref{fig:JFD-even-1} 
shows an example of the joint  frequency distribution. 

\section{Particle number estimation}\label{PNEstimation}

It is interesting to know how many particles can be produced by the gravitational perturbations around a black hole, e.g., how many particles are produced 
by the coalescence of two black holes in events of the type observed by LIGO. In this section, we provide a crude estimate of the total scalar particle number. 

Without loss of generality, we only consider the effect of odd-parity QNMs.
Using the decomposition theorem \cite{Law00}, the joint frequency distribution Eq. (\ref{JointDistribution}) can be decomposed as
\begin{eqnarray}
\mathcal{F}^{(o)}_{lml'}(\omega, \omega') = \sum_k r_k e^{i \varphi_k}\psi^*_{1k}(\omega) \psi^*_{2k}(\omega')
\end{eqnarray}
where $\{\psi_{1k}(\omega) \}$ and $\{\psi_{2k}(\omega') \}$ both form an orthonormal set of localized wave packet mode bases. The corresponding 
localized mode operators are 
\begin{eqnarray}
\hat{A}_{k l m} &=& \int \mathrm d \omega ~ \psi_{1k}(\omega) \hat{a}_{\omega l m}, \nonumber \\
\hat{B}_{k l' , -m} &=& \int \mathrm d \omega' ~ \psi_{2k}(\omega') \hat{a}_{\omega' l' , -m}.
\end{eqnarray}
The time evolution operator for given $l, m, l'$ becomes
\begin{eqnarray}
\hat{U}^{(o)}_{lml'} &\approx& \exp \bigg\{\sum_k r_k e^{i \varphi_k} \hat{A}^{\dag}_{k l m}  \hat{B}^{\dag}_{k l', -m} - \text{h.c.} \bigg\} \nonumber \\
&\approx& \bigotimes_k \exp \bigg\{r_k e^{i \varphi_k} \hat{A}^{\dag}_{k l m}  \hat{B}^{\dag}_{k l', -m} - \text{h.c.} \bigg\}.
\end{eqnarray}
We can see that $r_k$ is the two-mode squeezing factor and $\varphi_k$ is the squeezing angle,
\begin{eqnarray}
r_k e^{i \varphi_k} =  \int \mathrm d \omega \int \mathrm d \omega'  ~\mathcal{F}^{(o)}_{lml'}(\omega, \omega') \psi_{1k}(\omega) \psi_{2k}(\omega').
\end{eqnarray}
If the initial state of the scalar field is the Boulware vacuum, the particle number for $\hat{A}_{k l m}$ and $\hat{B}_{k l', -m}$ modes are both $\sinh^2(r_k)$. 
Therefore the total particle number is 
\begin{equation}
N = 2 \sum_{l m l'} \sum_k \sinh^2(r_k). 
\end{equation}

The maximum squeezing factor $r_{\text{max}}$ can be obtained by choosing $\psi_{1k}(\omega)$ and  $\psi_{2k}(\omega')$ such that they maximally 
overlap with the joint frequency distribution $\mathcal{F}^{(o)}_{lml'}(\omega, \omega')$. According to the behaviour of $\mathcal{F}^{(o)}_{lml'}(\omega, \omega')$, 
as shown by Fig. \ref{fig:JFD-1}, we can choose $\psi_{1k}(\omega)\psi_{2k}(\omega')$ as a two dimensional tophat function with width $\omega_I$ and length $\omega_R$, respectiely.
The maximum squeezing factor is approximately
\begin{eqnarray}\label{Max-r}
r_{\text{max}} &\sim& \sqrt{\omega_R \omega_I} \big| \mathcal{F}^{(o)}_{lml'}\big|_{\text{max}}
\end{eqnarray}
where $ \big | \mathcal{F}^{(o)}_{lml'}\big|_{\text{max}} $ is given by Eq. (\ref{MaxSqueezing}).

The amplitude of the ingoing gravitational waves on the horizon, $C_{\omega_0 l_0}$, remains unknown. It is related to the energy flux of the gravitational waves
falling into the black hole by 
\begin{eqnarray}\label{EnergyFlux}
\frac{\mathrm d E_{\text{hole}}}{\mathrm d t} = \frac{32 (2M)^4 \Omega_R^2(4\Omega_R^2+1)(\Omega_R^2+1)}{36\Omega_R^2 + l_0^2(l_0+1)^2(l_0^2+l_0-2)^2}
|C_{\omega_0 l_0}|^2,
\end{eqnarray}
This relation is obtained by making $a \rightarrow 0$ and integrating over all angles from a similar relation (Eq. (4.44) in \cite{Teukolsky74}) in the Kerr background spacetime. 
We assume that during time $\Delta t$, the amount of gravitational wave energy falling into the black hole is $\Delta E_{\text{hole}} \equiv \eta M_{\odot}$
(here $M_{\odot}$ is the solar mass, which is about $1.99 \times 10^{30}$ kg), then
\begin{eqnarray}\label{CEstimate}
|C_{\omega_0 l_0}| \sim  \sqrt{\frac{36\Omega_R^2 + l_0^2(l_0+1)^2(l_0^2+l_0-2)^2}{32 (2M)^4 \Omega_R^2(4\Omega_R^2+1)(\Omega_R^2+1)}}
\sqrt{\frac{\eta M_{\odot}}{\Delta t}}. \nonumber\\
\end{eqnarray}
As an example, we consider the fundamental QNM of a Schwarzschild black hole, for which $l_0 = 2, \Omega_R = 0.7474, \Omega_I = 0.178$. 
By using Eqs. (\ref{MaxSqueezing}), (\ref{Max-r}) and (\ref{CEstimate})
we find
\begin{eqnarray}
r_{\text{max}} &\sim& 0.57 \times \sqrt{\frac{\eta M_{\odot}}{\Delta t}} \sim 0.4 \sqrt{\eta} \times \bigg( \frac{10^{-5} s}{\Delta t} \bigg)^{1/2}. 
\end{eqnarray}
For the first gravitational wave event, GW150914, detected by LIGO \cite{GW16:1}, the ring-down time is about $0.01$ second, so 
\begin{eqnarray}
r_{\text{max}} \sim \sqrt{\eta} \times 10^{-2}. 
\end{eqnarray}
We note that even if $\eta$ is at the order of one (one solar mass of gravitational wave energy falls into the black hole), 
the particle number created by the ring-down is very small. This is due to the fact that at the ring-down stage, the black hole oscillation
dies away very quickly. 

One important thing to note, from Eqs. (\ref{MaxSqueezing}), (\ref{Max-r}) and (\ref{CEstimate}), is that $r_{\text{max}} \propto \Omega_I^{-5/2}$, where
$1/\Omega_I$ characterizes the time scale of the black hole oscillation. If this is qualitatively valid at the inspiral and merger stages, one might expect
the particle generation is quite significant because the oscillation of the black hole lasts much longer than that of the ring-down stage. For example, if 
the time scale is 30 times longer, $r_{\text{max}} \sim 50$ and the total particle number $\sim 10^{42}$. However, our calculation is only valid at the ring-down 
stage and further studies need to be done for the inspiral and merger stages.

\section{Conclusion}\label{conclusion}

We studied the coupling between the gravitational QNMs of a Schwarzschild black hole and a massless scalar field, and showed that
scalar particles can be produced by the gravitational perturbations. This is contrary to the plane 
gravitational wave case where particle creation is forbidden due to the violation of momentum conservation. In the Schwarzschild black hole case, 
the total angular momentum of the QNMs and the produced pair of particles is conserved. In arriving at the above conclusions, 
we explicitly derived the interaction Hamiltonian for the scalar field which shows that the QNMs play the role as a multimode squeezer. 
The QNMs squeeze the initial state of the scalar field and produce particles. If the initial state of the scalar field
is a vacuum state (Boulware vacuum), then the final state is a squeezed vacuum state, indicating that the QNMs amplify the vacuum 
fluctuations and create particles. If initially there exists  Hawking radiation (a thermal state with its temperature proportional to the
surface gravity of the black hole) the QNMs  squeeze the Hawking radiation and amplify it. 
Interestingly, the presence of gravitational perturbations results in coupling between different Hawking particles, and therefore may build correlations between
them, modifying the thermal characteristic. Modification of the Unruh radiation via  modulation of the Rindler horizon, mimicked by non-uniform acceleration, was recently studied \cite{Kempf17}. 
In realistic astrophysical situations, the CMB temperature  is higher than the Hawking temperature of astrophysical black holes, so the amplification of the CMB around a black  hole by the QNMs will be more significant than the amplification of  Hawking radiation. 

How significant  the amplification is depends on the squeezing amplitude. We showed that the squeezing amplitude, Eq. (\ref{MaxSqueezing}), is proportional to the 
amplitude of the QNMs, which is reasonable because larger gravitational perturbations would create more particles. 
In addition, the maximal squeezing amplitude is proportional to the cube of the black hole mass and the real part of the QNMs frequency, 
and is inversely proportional to the cube of the imaginary part of the QNM frequency. This implies that for a given amplitude of the QNMs a larger
black hole would create more particles. Furthermore, the particle creation efficiency is higher for lower decaying QNMs. For Schwarzschild
black holes, the damping of the QNMs is fast. The least damped mode has $2M \omega_I = 0.178$. For extreme Kerr black holes, there exists
QNMs with very small damping rate, called Zero-Damping modes \cite{Yang13}. We expect that our result is qualitatively correct for the Kerr black holes, which
implies the particle creation by gravitational perturbations around an extreme Kerr black hole is much more efficient than that around a Schwarzschild black hole. 

%
%


\section*{ACKNOWLEGEMENTS}
 
We would like to thank Jolien Creighton for helpful correspondence.
This research was supported in part by Australian Research Council Centre of Excellence of Quantum 
Computation and Communication Technology (Project No. CE110001027),
and in part by the Natural Sciences and Engineering Research Council of Canada.

\appendix
\vspace{0.5cm}
\section{Radial Integrals 
\label{appendixA}}
It is difficult to find exactly analytic results for the radial integrals 
because we do not have analytic
solutions for ${}_{-2}R_{\omega_0 l_0}$ and $R_{\omega l}$. However, it is possible to find approximately analytic results by using
the asymptotic behaviour of ${}_{-2}R_{\omega_0 l_0}$ and $R_{\omega l}$.  When $r_* \rightarrow +\infty$ (spatial infinity), 
the integrands of the radial integrals 
are both proportional to $e^{i (\omega_0 - \omega - \omega') r_*}$ according to Eqs. (\ref{boundary:QNM}) 
and (\ref{upasymptotic}); when $r_* \rightarrow -\infty$ (event horizon), they are proportional to $e^{- i (\omega_0 - \omega - \omega') r_*}$. 
The QNM frequency is a complex number, $\omega_0 = \omega_R - i \omega_I$, so 
the integrands are proportional to $e^{\omega_I r_*}$ when $r_* \rightarrow +\infty$, and $e^{-\omega_I r_*}$ when $r_* \rightarrow -\infty$. 
Since $\omega_I > 0$, the integrands are divergent at the spatial infinity and on the event horizon, which implies the radial integrals are not well defined. 
This formal divergence can be resolved by ultilizing the analyticity of the integrands \cite{Leaver86}: they are analytic in the complex $r$ plane except at two points
$r = 2M$ and $r = \infty$. 
A contour which encloses these two branch points $r = 2M$ and $r = \infty$ is chosen by Leaver \cite{Leaver86} such that the integral along it is well defined and finite. 
Sun and Price \cite{SunPrice88} discussed in detail how to construct Leaver's contour by analytic continuation and restored a factor, $1/(1 - e^{2\pi \Omega_0})$,
 that is missed in \cite{Leaver86}.

\begin{figure}[ht!]
\centering
\begin{tikzpicture}[scale=1.3]

\begin{scope}[color=black!20]
 \fill (1.65, -0.35) -- (2.35, -0.35) -- (2.35, 1.0) -- (1.65, 1.0) -- cycle; 
 \fill (1.65, 2) -- (2.35, 2) -- (2.35, 3.2) -- (1.65, 3.2) -- cycle; 
\end{scope}

\begin{scope}[->, line width=0.6pt]
  \draw (-0.5, 0) -- (xyz cs:x=4);
  \draw (0, -0.5) -- (xyz cs:y=3);
\end{scope}

\fill (2, 0) circle (2pt);
\fill (0, 0) circle (0.8pt);

\begin{scope}[line width=1pt]
 \draw (1.85, 0.26) arc [start angle=120, end angle=420, radius=0.3cm];
 \draw (1.85, 3) -- (1.85, 0.25); 
 \draw (2.15, 0.25) -- (2.15, 3);
\end{scope}

\begin{scope}[->, line width=1pt]
 \draw (2.15, 0.26) arc [start angle=60, end angle=-95, radius=0.3cm];
 \draw (1.85, 0.25) -- (1.85, 1.5); 
 \draw (2.15, 3) -- (2.15, 1.4);
\end{scope}


\draw [ decorate, decoration=zigzag] (2, 0) -- (2, 3);

\begin{scope}
 \draw (4.5, 0) node {Re($r$)};
 \draw (0, 3.3) node {Im($r$)};
 \draw (2.1, -0.6) node {$r=2M$};
 \draw (2.85, 1.7) node {$C$};
 \draw (-0.2, -0.2) node {$O$};
 
 \draw (3, 2.7) node {${\it out}$ region};
 \draw (3, 0.3) node {${\it in}$ region};
\end{scope}

\path [->] (2.7, 1.7) edge (2.2, 1.7); 

\end{tikzpicture}

\caption{\footnotesize The contour $C$ and branch cut when $\omega_R - \omega - \omega' > 0$. The two shaded regions are referred to as 
${\it in}$ (close to the horizon) and ${\it out}$ (around $r = \infty$) regions, respectively. }
\label{fig:Contour}
\end{figure}
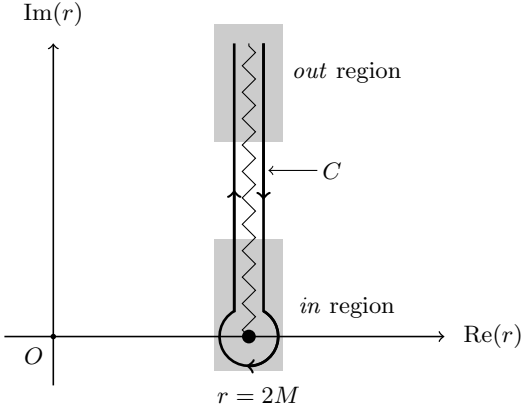

Noting that $r = 2M$ and $r = \infty$ are two branch
points, the branch cut can be chosen as a line perpendicular to the real $r$ axis, starting at $r = 2M$ and ending at $r = \infty$.  It lies in the upper 
complex $r$ plane if $\omega_R - \omega - \omega' > 0$, as shown in Fig. \ref{fig:Contour}, and in the lower complex $r$ plane if $\omega_R - \omega - \omega' < 0$. When 
$\omega_R - \omega - \omega' > 0$, the contour $C$ begins at $r = \infty$, right next to the branch cut, moves downward to $r = 2M$, where it wraps
and, left next to the branch cut, moves upward to $r = \infty$, as shown in Fig. \ref{fig:Contour}. We refer to the region near $r = 2M$ as the ${\it in}$ region and
the region around $r = \infty$ as the ${\it out}$ region, as schematically represented by the shaded region in Fig. \ref{fig:Contour}. 
By analytically extending the integrands to the complex $r$ plane we see that along
the contour $C$ the integrands exponentially decay in the ${\it out}$ region, which thus remove the formal divergence. In addition, the exponential decay
of the integrands in the ${\it out}$ region implies that the main contributions to the integrals are from the ${\it in}$ region. 


We describe in detail how to find the approximately analytic result for the radial integral $\mathcal{I}_{r1}$; the result for $\mathcal{I}_{r2}$ and $\mathcal{I}_{r3}$ can 
be obtained in a similar way. At spatial infinity ($r \rightarrow \infty $), by using Eqs. (\ref{boundary:QNM}) and (\ref{upasymptotic}), the integrand of $\mathcal{I}_{r1}$ can be 
approximated as 
\begin{eqnarray}\label{Inte-1-infinity}
&&\frac{1}{f} \bigg[ \bigg(\frac{\mathrm d}{\mathrm d r_*}-i\omega_0 - \frac{2 f}{r}  \bigg) {}_{-2}R_{\omega_0 l_0}  \bigg]
\frac{\mathrm d}{\mathrm d r} \bigg(\frac{R^*_{\omega l}}{r} \bigg) \bigg(\frac{R^*_{\omega' l'}}{r} \bigg) \nonumber \\
&\approx& - B^*_{\omega l} B^*_{\omega' l'} D_{\omega_0 l_0} \bigg (\frac{i \omega}{f} + \frac{1}{r} \bigg) e^{i (\omega_0 - \omega - \omega') r_*} \nonumber \\
&\approx& - \frac{B^*_{\omega l} B^*_{\omega' l'} D_{\omega_0 l_0}}{2M}   \big( i \Omega x^{i\Delta} +x^{-1+i\Delta} \big) e^{i\Delta x}, 
\end{eqnarray}
where we have defined a dimensionless radial coordinate $x=r/2M$.
%
%
Near the event horizon ($r \rightarrow 2M$), according to Eqs. (\ref{boundary:QNM}) and (\ref{upasymptotic}), 
 the integrand of $\mathcal{I}_{r1}$ can be approximated as 
\begin{eqnarray}\label{Inte-1-horizon}
&&\frac{1}{f} \bigg[ \bigg(\frac{\mathrm d}{\mathrm d r_*}-i\omega_0 - \frac{2 f}{r}  \bigg) {}_{-2}R_{\omega_0 l_0}  \bigg]
\frac{\mathrm d}{\mathrm d r} \bigg(\frac{R^*_{\omega l}}{r} \bigg) \bigg(\frac{R^*_{\omega' l'}}{r} \bigg) \nonumber \\
&\approx& 2 A^*_{\omega l} A^*_{\omega' l'} C_{\omega_0 l_0} \bigg [ x^{-1} (1 - i \Omega_0 x)(i \Omega x -1) (x-1)^{1 - i\Delta} \nonumber \\
&& + i \Omega (1 - i \Omega_0 x) (x - 1)^{-i \Delta}  \bigg] e^{- i\Delta x}, 
\end{eqnarray}
where we have only kept the term proportional to $e^{-i (\omega_0 - \omega - \omega') r_*}$ owing to the rotating wave approximation. 
As discussed before, the integration along the real $r$ axis is formally divergent. The integrands Eqs. (\ref{Inte-1-infinity}) and (\ref{Inte-1-horizon})
are analytically extended to the whole complex $r$ plane. Along the contour $C$ in the ${\it out}$ region, $e^{i\Delta x} \sim e^{-2M\omega_I |x|}$,
which means the integrand of $\mathcal{I}_{r1}$ exponentially decays. We therefore expect that the integral $\mathcal{I}_{r1}$ 
is finite along the contour $C$. Unfortunately, we cannot find an analytic expression for the integrand on the whole contour $C$. Numerical
techniques need to be introduced to perform the contour integration. However, it may be possible that an approximate result can be obtained
by using only the asymptotic behaviour of the integrand. Note that in the ${\it out}$ region the integrand (\ref{Inte-1-infinity}) exponentially decays 
and contributes very little to the total integral. 
Introducing another exponential decaying function in the ${\it out}$ region would not introduce large deviation
to the integral. We therefore replace Eq. (\ref{Inte-1-infinity}) by Eq. (\ref{Inte-1-horizon}) with the factor $e^{- i\Delta x}$ replaced by $e^{i\Delta x}$.
In the {\it in} region, the asymptotic expression for the integrand is Eq. (\ref{Inte-1-horizon}) which dominates the contribution to the integral. 
In the limit of $i \Delta \sim 0$, which is the case that we are mostly interested in, $e^{- i\Delta x} \approx e^{i\Delta x}$. We thus replace $e^{- i\Delta x}$
by $e^{i\Delta x}$ in Eq. (\ref{Inte-1-horizon}) and get an approximately asymptotic expression. In summary, we approximate the original integrand by
\begin{eqnarray}\label{Inte-1-approximation}
&&\frac{1}{f} \bigg[ \bigg(\frac{\mathrm d}{\mathrm d r_*}-i\omega_0 - \frac{2 f}{r}  \bigg) {}_{-2}R_{\omega_0 l_0}  \bigg]
\frac{\mathrm d}{\mathrm d r} \bigg(\frac{R^*_{\omega l}}{r} \bigg) \bigg(\frac{R^*_{\omega' l'}}{r} \bigg) \nonumber \\
&\approx&2 A^*_{\omega l} A^*_{\omega' l'}  C_{\omega_0 l_0} \bigg [ x^{-1} (1 - i \Omega_0 x)(i \Omega x -1) (x-1)^{1 - i\Delta} \nonumber \\
&& + i \Omega (1 - i \Omega_0 x) (x - 1)^{-i \Delta}  \bigg] e^{i\Delta x}
\end{eqnarray}
along the whole contour $C$. Obviously, this is a very crude approximation since we have ignored the behaviour of the integrand in the intermediate region.
The validity of this approximation has to be verified by numerical calculation. However, we expect that this approximation provides a lower bound for 
the exact integral since near the horizon we replace an exponentially growing function by an an exponentially decaying function. The advantage of this
approximation is that we can obtain an analytic result for the radial integral $\mathcal{I}_{r1}$. 

From Eq. (\ref{Inte-1-approximation}) we see that basically
we need to calculate
\begin{equation}
\oint_{C} \mathrm d x ~ e^{i \Delta x} (x-1)^{n_1 - i \Delta} x^{n_2}
\end{equation}
where $n_1, n_2$ are two integers, $C$ is the contour we introduced, as shown in Fig. \ref{fig:Contour}. Defining a new variable $u = i \Delta (x -1)$,  we find \cite{Morse53}
\begin{eqnarray}\label{ContourIntegration}
&&\oint_{C} \mathrm d x ~ e^{i \Delta x} (x-1)^{n_1 - i \Delta} x^{n_2} \nonumber \\
&=& (i \Delta)^{i\Delta - n_1 - 1} e^{i\Delta} \oint_{F} \mathrm d u ~ e^u u^{n_1 - i \Delta} \bigg(1 + \frac{u}{i \Delta} \bigg)^{n_2} \nonumber \\
&=& \frac{2 \pi i (-1)^{n_1 + 1 - i \Delta} e^{i \Delta}}{\Gamma(-n_1 + i \Delta)} \nonumber\\
&&\times ~U(n_1 + 1 - i \Delta, n_1 + n_2 + 2 - i \Delta, - i \Delta),
\end{eqnarray}
where $F$ is the contour illustrated by Morse and Feshbach's Fig. 5.1.2 \cite{Morse53} (page 612), $\Gamma(z)$ is the Gamma's function and 
$U(a, c, z)$ is the confluent hypergeometric function \cite{Abramowitz} (Note that the function $U_2(a, c, z)$ defined by Morse and Feshbach is related to 
$U(a, c, z)$ by $U_2(a, c, z) = e^{i\pi a } U(a, c, z)$). Therefore the radial integral $\mathcal{I}_{r1}$ can be approximated as
\begin{widetext}
\begin{eqnarray}\label{Integral-1-approximation}
\mathcal{I}_{r1} &\approx& - \frac{8 \pi M i \sqrt{D_0}}{\sqrt{2l_0(l_0+1)}} \frac{A^*_{\omega l} A^*_{\omega' l'} C_{\omega_0 l_0}e^{i \Delta}}{1 - e^{2\pi \Omega_0}}
\bigg\{ \frac{(-1)^{2-i\Delta}}{\Gamma(-1 + i \Delta)} \bigg[ \Omega \Omega_0 ~ U(2-i\Delta, 4-i\Delta, -i\Delta) + i (\Omega + \Omega_0) ~ U(2-i\Delta, 3-i\Delta, -i\Delta) \nonumber \\
&& - U(2-i\Delta, 2-i\Delta, -i\Delta) \bigg] + \frac{(-1)^{1-i\Delta}}{\Gamma(i \Delta)} \bigg[ \Omega \Omega_0 ~ U(1-i\Delta, 3-i\Delta, -i\Delta)
+ i \Omega ~ U(1-i\Delta, 2-i\Delta, -i\Delta) \bigg] \bigg\} \nonumber \\
&\approx& - \frac{8 \pi M i \sqrt{D_0}}{\sqrt{2l_0(l_0+1)}} \frac{A^*_{\omega l} A^*_{\omega' l'} C_{\omega_0 l_0} e^{i \Delta}}{1 - e^{2\pi \Omega_0}}
\bigg\{ \frac{(i \Delta)^{i\Delta - 2}}{\Gamma(-1 + i \Delta)} \bigg[ -\frac{2\Omega \Omega_0}{i\Delta} + 2\Omega \Omega_0 + i (\Omega + \Omega_0)\bigg] \nonumber \\
 &&- \frac{(-1)^{2-i\Delta}}{\Gamma(-1 + i \Delta)} U(2-i\Delta, 2-i\Delta, -i\Delta)
 + \frac{(i\Delta)^{i\Delta-1}}{\Gamma(i \Delta)} \bigg( -\frac{\Omega \Omega_0}{i\Delta}  + 2\Omega \Omega_0 + i \Omega \bigg) \bigg\},
\end{eqnarray}
\end{widetext}
where we have used the fact that \cite{NIST}
\begin{equation}
U(a, a+n+1, z) = \frac{z^{-a}}{\Gamma(a)} \sum_{k=0}^{n} \dbinom{n}{k} \Gamma(a+k) z^{-k},
\end{equation}
with $n = 0, 1, 2, \cdots$.
In the limit of $i\Delta \sim 0$, from Eq. (\ref{Integral-1-approximation}), we obtain the dominant term
\begin{equation}
\mathcal{I}_{r1} \approx - \frac{\sqrt{D_0}}{\sqrt{2l_0(l_0+1)}} \frac{16 \pi i M \Omega \Omega_0 }{(i \Delta)^2} 
\frac{A^*_{\omega l} A^*_{\omega' l'}  C_{\omega_0 l_0}}{1 - e^{2\pi \Omega_0}}. 
\end{equation}
%
%

The calculation of $\mathcal{I}_{r2}$ is very similar to that of $\mathcal{I}_{r1}$. At spatial infinity ($r \rightarrow \infty$), According to Eqs. (\ref{boundary:QNM}) and (\ref{upasymptotic}), 
the integrand of $\mathcal{I}_{r2}$ can be approximated as
\begin{widetext}
\begin{eqnarray}\label{Inte-2-infinity}
&&\frac{1}{r^2 f^2}
\bigg[(i \omega_0 r^2 - M)\frac{\mathrm d}{\mathrm d r_*}{}_{-2}R_{\omega_0 l_0} - \bigg(\frac{1}{2} \mu_0^2 f - i\omega_0 (-3r+7M) - \omega_0^2 r^2  \bigg){}_{-2}R_{\omega_0 l_0}\bigg]
\bigg(\frac{R^*_{\omega l}}{r} \bigg) \bigg(\frac{R^*_{\omega' l'}}{r} \bigg) \nonumber \\
&\approx& B^*_{\omega l} B^*_{\omega' l'} D_{\omega_0 l_0}\bigg[ \frac{3 i \omega_0}{f}\bigg(1-\frac{1}{f} \bigg) + \frac{6i M \omega_0}{r f^2} - \frac{\mu_0^2}{2 rf} - \frac{3M}{r^2 f} \bigg]
e^{i(\omega_0 - \omega - \omega') r_*} \nonumber \\
&\approx& -\frac{1}{4M} B^*_{\omega l} B^*_{\omega' l'} D_{\omega_0 l_0} \big[  \mu_0^2 (x-1)^{-1+i\Delta} + 3 x^{-1} (x-1)^{-1+i\Delta}\big] e^{i\Delta x}. 
\end{eqnarray}
\end{widetext}
The dominant term is $ e^{i\Delta x}/x$ when $x$ is large. Near the event horizon ($r \rightarrow 2M$). According to Eqs. (\ref{QNhorizon}) and (\ref{upasymptotic}),  the integrand of $\mathcal{I}_{r2}$ can be approximated as
\begin{widetext}
\begin{eqnarray}\label{Inte-2-horizon}
&&\frac{1}{r^2 f^2}
\bigg[(i \omega_0 r^2 - M)\frac{\mathrm d}{\mathrm d r_*}{}_{-2}R_{\omega_0 l_0} - \bigg(\frac{1}{2} \mu_0^2 f - i\omega_0 (-3r+7M) - \omega_0^2 r^2  \bigg){}_{-2}R_{\omega_0 l_0}\bigg]
\bigg(\frac{R^*_{\omega l}}{r} \bigg) \bigg(\frac{R^*_{\omega' l'}}{r} \bigg) \nonumber \\
&\approx& A^*_{\omega l} A^*_{\omega' l'} C_{\omega_0 l_0} \bigg[ 2\omega_0^2 r^2 + i \omega_0 r + 4i M \omega_0 - \frac{1}{2}\mu_0^2 f - \frac{4M}{r} + \frac{4M^2}{r^2} \bigg]
e^{-i(\omega_0 - \omega - \omega') r_*} \nonumber \\
&\approx& A^*_{\omega l} A^*_{\omega' l'} C_{\omega_0 l_0} \bigg[ (2 \Omega_0^2 x^2 + i \Omega_0 x + 2i \Omega_0 - 2 x^{-1} + x^{-2}) (x-1)^{-i\Delta} 
- \frac{1}{2} \mu_0^2 x^{-1} (x-1)^{1-i\Delta}\bigg] e^{-i\Delta x}. 
\end{eqnarray}
By analytically extending Eqs. (\ref{Inte-2-infinity}) and (\ref{Inte-2-horizon}) to the complex $r$ plane and 
using the same approximation that leads to Eq. (\ref{Inte-1-approximation}), we obtain an approximate expression for the integrand of $\mathcal{I}_{r2}$
along the whole contour $C$,
\begin{eqnarray}\label{Inte-2-approximation}
&&\frac{1}{r^2 f^2}
\bigg[(i \omega_0 r^2 - M)\frac{\mathrm d}{\mathrm d r_*}{}_{-2}R_{\omega_0 l_0} - \bigg(\frac{1}{2} \mu_0^2 f - i\omega_0 (-3r+7M) - \omega_0^2 r^2  \bigg){}_{-2}R_{\omega_0 l_0}\bigg]
\bigg(\frac{R^*_{\omega l}}{r} \bigg) \bigg(\frac{R^*_{\omega' l'}}{r} \bigg) \nonumber \\
&\approx& A^*_{\omega l} A^*_{\omega' l'} C_{\omega_0 l_0} \bigg[ (2 \Omega_0^2 x^2 + i \Omega_0 x + 2i \Omega_0 - 2 x^{-1} + x^{-2}) (x-1)^{-i\Delta} 
- \frac{1}{2} \mu_0^2 x^{-1} (x-1)^{1-i\Delta}\bigg] e^{i\Delta x}. 
\end{eqnarray}
\end{widetext}
We can see that in the {\it out} region on the contour $C$, the dominant term in Eq. (\ref{Inte-2-approximation}) is $x^2 e^{i\Delta x}$ which is 
greater than that in Eq. (\ref{Inte-2-infinity}), $ e^{i\Delta x}/x$. However, these two terms are both exponentially suppressed so that their contribution
to the total integration is small.  

We therefore expect that this approximation only introduces a small error. The main contribution to the integration comes 
from the {\it in} region where $x$ is not large. In the limit of $i\Delta \sim 0$, which is the case that we are mostly interested in, $e^{-i\Delta x} \approx e^{i\Delta x}$.
We therefore expect that Eq. (\ref{Inte-2-approximation}) is a good approximation to Eq. (\ref{Inte-2-horizon}) in the {\it in} region. Note that we replace an exponential
growing function by an exponentially decaying function in the {\it in} region; the final result provides a lower bound for the exact radial integral $\mathcal{I}_{r2}$. 
Using Eq. (\ref{ContourIntegration}) we have 
\begin{widetext}
\begin{eqnarray}\label{Integral-2-approximation}
\mathcal{I}_{r2} &\approx& 4 \pi M i \frac{A^*_{\omega l} A^*_{\omega' l'} C_{\omega_0 l_0}e^{i \Delta}}{1 - e^{2\pi \Omega_0}}
\bigg\{ \frac{(-1)^{1-i\Delta}}{\Gamma(i \Delta)} \bigg[ 2\Omega_0^2 U(1-i\Delta, 4-i\Delta, -i\Delta) + i \Omega_0 U(1-i\Delta, 3-i\Delta, -i\Delta)  \nonumber \\
&&+ 2i \Omega_0 U(1-i\Delta, 2-i\Delta, -i\Delta) - 2 U(1-i\Delta, 1-i\Delta, -i\Delta) + U(1-i\Delta, -i\Delta, -i\Delta) \bigg] \nonumber \\
&&- \frac{(-1)^{2-i\Delta}}{2 \Gamma(-1 + i \Delta)} \mu_0^2 U(2-i\Delta, 2-i\Delta, -i\Delta) \bigg\} \nonumber \\
&\approx&4 \pi M i  \frac{A^*_{\omega l} A^*_{\omega' l'} C_{\omega_0 l_0}e^{i \Delta}}{1 - e^{2\pi \Omega_0}}
\bigg\{ \frac{(i\Delta)^{i\Delta-1}}{\Gamma(i \Delta)} \bigg[ \frac{4\Omega_0^2}{(i\Delta)^2}  - \frac{i \Omega_0 + 10 \Omega_0^2}{i\Delta} 
+ 8 \Omega_0^2 + 4i \Omega_0 \bigg] \nonumber\\
&&- \frac{(-1)^{1-i\Delta}}{\Gamma(i \Delta)}\bigg[2 U(1-i\Delta, 1-i\Delta, -i\Delta) - U(1-i\Delta, -i\Delta, -i\Delta) \bigg]
- \frac{(-1)^{2-i\Delta}}{2 \Gamma(-1 + i \Delta)} \mu_0^2 U(2-i\Delta, 2-i\Delta, -i\Delta) \bigg\}. \nonumber \\
\end{eqnarray}
\end{widetext}
The dominant term in the limit of $i\Delta \sim 0$ is 
\begin{equation}
\mathcal{I}_{r2} \approx \frac{16 \pi i M \Omega_0^2 }{(i \Delta)^2}
\frac{A^*_{\omega l} A^*_{\omega' l'} C_{\omega_0 l_0}}{1 - e^{2\pi \Omega_0}}. 
\end{equation}

At spatial infinity ($r \rightarrow \infty$), the integrand of $\mathcal{I}_{r3}$ can be approximated as 
\begin{eqnarray}\label{Inte-3-infinity}
&&\frac{1}{f^2} {}_{-2}R_{\omega_0 l_0} \bigg(\frac{R^*_{\omega l}}{r} \bigg) \bigg(\frac{R^*_{\omega' l'}}{r} \bigg) \nonumber \\
&\approx& 2M B^*_{\omega l} B^*_{\omega' l'} D_{\omega_0 l_0} x^3 (x-1)^{-2+i\Delta} e^{i\Delta x}.
\end{eqnarray}
The dominant term is $x e^{i\Delta x}$ when $x$ is large. Near the event horizon ($r \rightarrow 2M$), 
the the integrand of $\mathcal{I}_{r3}$ can be approximated as
\begin{eqnarray}\label{Inte-3-horizon}
&&\frac{1}{f^2} {}_{-2}R_{\omega_0 l_0} \bigg(\frac{R^*_{\omega l}}{r} \bigg) \bigg(\frac{R^*_{\omega' l'}}{r} \bigg) \nonumber \\
&\approx& 4M^2 A^*_{\omega l} A^*_{\omega' l'} C_{\omega_0 l_0} x^2 (x-1)^{-i\Delta} e^{-i\Delta x}.
\end{eqnarray}
By analytically extending Eqs. (\ref{Inte-3-infinity}) and (\ref{Inte-3-horizon}) to the complex $r$ plane and
using the same approximation as before, we obtain an approximate expression for the integrand of $\mathcal{I}_{r3}$
along the whole contour $C$,
\begin{eqnarray}
&&\frac{1}{f^2} {}_{-2}R_{\omega_0 l_0} \bigg(\frac{R^*_{\omega l}}{r} \bigg) \bigg(\frac{R^*_{\omega' l'}}{r} \bigg) \nonumber \\
&\approx& 4M^2 A^*_{\omega l} A^*_{\omega' l'} C_{\omega_0 l_0} x^2 (x-1)^{-i\Delta} e^{i\Delta x}.
\end{eqnarray}
Using Eq. (\ref{ContourIntegration}) we have
\begin{eqnarray}
\mathcal{I}_{r3} &\approx& -4\pi i \sqrt{D_0}M \Omega \Omega' \frac{A^*_{\omega l} A^*_{\omega' l'} C_{\omega_0 l_0}e^{i \Delta}}{1 - e^{2\pi \Omega_0}} \frac{(-1)^{1-i\Delta} }{\Gamma(i\Delta)} \nonumber\\
&&\times ~U(1-i\Delta, 4-i\Delta,-i\Delta) \nonumber\\
&\approx&  -4\pi i \sqrt{D_0}M \Omega \Omega' \frac{A^*_{\omega l} A^*_{\omega' l'} C_{\omega_0 l_0}e^{i \Delta}}{1 - e^{2\pi \Omega_0}} \frac{(i\Delta)^{i\Delta-1} }{\Gamma(i\Delta)} \nonumber\\
&&\times \bigg [4 - \frac{5}{i\Delta} + \frac{2}{(i\Delta)^2} \bigg]. 
\end{eqnarray}
The dominant term in the limit of $i\Delta \sim 0$ is
\begin{equation}
\mathcal{I}_{r3} \approx - \frac{8 \pi i \sqrt{D_0} M \Omega \Omega' }{(i \Delta)^2} \frac{A^*_{\omega l} A^*_{\omega' l'} C_{\omega_0 l_0}}{1 - e^{2\pi \Omega_0}}. 
\end{equation}

At spatial infinity ($r \rightarrow \infty$), the integrand of $\mathcal{I}_{r4}$ can be approximated as 
\begin{eqnarray}\label{Inte-4-infinity}
&&{}_{-2}R_{\omega_0 l_0} \frac{\mathrm d}{\mathrm d r} \bigg(\frac{R^*_{\omega l}}{r} \bigg) \frac{\mathrm d}{\mathrm d r} \bigg(\frac{R^*_{\omega' l'}}{r} \bigg) \nonumber \\
&\approx& \frac{1}{2M} B^*_{\omega l} B^*_{\omega' l'} D_{\omega_0 l_0} \big[-\Omega \Omega' x^3 (x-1)^{-2} + x^{-1}\nonumber\\
&&+ i(\Omega+\Omega') x (x-1)^{-1}  \big] (x-1)^{i\Delta}  e^{i\Delta x}.
\end{eqnarray}
Near the event horizon ($r \rightarrow 2M$), the the integrand of $\mathcal{I}_{r4}$ can be approximated as
\begin{eqnarray}\label{Inte-4-horizon}
&&{}_{-2}R_{\omega_0 l_0} \frac{\mathrm d}{\mathrm d r} \bigg(\frac{R^*_{\omega l}}{r} \bigg) \frac{\mathrm d}{\mathrm d r} \bigg(\frac{R^*_{\omega' l'}}{r} \bigg) \nonumber \\
&\approx& -A^*_{\omega l} A^*_{\omega' l'} C_{\omega_0 l_0} \big[\Omega \Omega' x^2 + i(\Omega+\Omega') (x-1)  \nonumber\\
&&- x^{-2} (x-1)^2 \big] (x-1)^{- i\Delta}  e^{-i\Delta x}.
\end{eqnarray}
By analytically extending Eqs. (\ref{Inte-4-infinity}) and (\ref{Inte-4-horizon}) to the complex $r$ plane and
using the same approximation as before, we obtain an approximate expression for the integrand of $\mathcal{I}_{r3}$
along the whole contour $C$,
\begin{eqnarray}\label{Inte-4-C}
&&{}_{-2}R_{\omega_0 l_0} \frac{\mathrm d}{\mathrm d r} \bigg(\frac{R^*_{\omega l}}{r} \bigg) \frac{\mathrm d}{\mathrm d r} \bigg(\frac{R^*_{\omega' l'}}{r} \bigg) \nonumber \\
&\approx& -A^*_{\omega l} A^*_{\omega' l'} C_{\omega_0 l_0} \big[\Omega \Omega' x^2 + i(\Omega+\Omega') (x-1)  \nonumber\\
&&- x^{-2} (x-1)^2 \big] (x-1)^{- i\Delta}  e^{i\Delta x}.
\end{eqnarray}
Using Eq. (\ref{ContourIntegration}) we have
\begin{widetext}
\begin{eqnarray}\label{Integral-4-approximation}
\mathcal{I}_{r4} &\approx& 4 \pi M i \sqrt{D_0} \frac{A^*_{\omega l} A^*_{\omega' l'} C_{\omega_0 l_0}e^{i \Delta}}{1 - e^{2\pi \Omega_0}}
\bigg\{ \frac{(-1)^{1-i\Delta}}{\Gamma(i \Delta)} \Omega \Omega' U(1-i\Delta, 4-i\Delta, -i\Delta) \nonumber\\
&&+ \frac{(-1)^{2-i\Delta}}{\Gamma(-1+ i \Delta)} i (\Omega + \Omega') U(2-i\Delta, 2-i\Delta, -i\Delta) 
- \frac{(-1)^{3-i\Delta}}{\Gamma(-2 + i \Delta)} U(3-i\Delta, 2-i\Delta, -i\Delta) \nonumber \\
&\approx& 4 \pi M i \sqrt{D_0} \frac{A^*_{\omega l} A^*_{\omega' l'} C_{\omega_0 l_0}e^{i \Delta}}{1 - e^{2\pi \Omega_0}}
\bigg\{ \frac{(i\Delta)^{i\Delta-1}}{\Gamma(i \Delta)} \Omega \Omega' \bigg [4 - \frac{5}{i\Delta} + \frac{2}{(i\Delta)^2} \bigg] \nonumber\\
&&+ \frac{(-1)^{2-i\Delta}}{\Gamma(-1+ i \Delta)} i (\Omega + \Omega') U(2-i\Delta, 2-i\Delta, -i\Delta) 
- \frac{(-1)^{3-i\Delta}}{\Gamma(-2 + i \Delta)} U(3-i\Delta, 2-i\Delta, -i\Delta).
\end{eqnarray}
\end{widetext}
The dominant term in the limit of $i\Delta \sim 0$ is
\begin{equation}
\mathcal{I}_{r4} \approx \frac{8 \pi i \sqrt{D_0} M \Omega \Omega' }{(i \Delta)^2} \frac{A^*_{\omega l} A^*_{\omega' l'} C_{\omega_0 l_0}}{1 - e^{2\pi \Omega_0}}. 
\end{equation}

\section{Explicit Expressions for Spin-weighted spherical harmonics \label{appendix:SWS}}

In order to calculate the the angular integrals, 
one needs to know the explicit expressions for the spin weighted
spherical harmonics ${}_{\pm 1} Y_{lm}$ and ${}_{\pm 2} Y_{lm}$. The spin weighted spherical harmonics ${}_s Y_{lm}$ for integers $s, l, m$ is defined from the spherical 
harmonics $Y_{lm}$ \cite{Goldberg67}, 
\begin{eqnarray}\label{SWH-definition}
{}_s Y_{lm} =
\begin{cases}
\sqrt{\frac{(l-s)!}{(l+s)!}} ~\eth^s Y_{lm}, &  0 \le s \le l ;  \\ 
\sqrt{\frac{(l+s)!}{(l-s)!}} (-1)^{s}~\bar{\eth}^{-s} Y_{lm}, & -l \le s \le 0,
\end{cases}
\end{eqnarray}
where $\eth$ and $\bar{\eth}$ are the spin-raising and spin-lowering operators, respectively. Assume that $\eta$ is a quantity of spin-weight $s$, then $\eth \eta$ is a quantity
of spin-weight $s+1$,
\begin{equation}\label{raisingspin}
\eth \eta \equiv -(\sin \theta)^s \bigg(\frac{\partial}{\partial \theta} + \frac{i}{\sin \theta} \frac{\partial}{\partial \phi} \bigg) \big[(\sin \theta)^{-s} \eta \big];
\end{equation}
and $\bar{\eth} \eta$ is a quantity of spin-weight $s-1$,
\begin{equation}\label{loweringspin}
\bar{\eth} \eta \equiv -(\sin \theta)^{-s} \bigg(\frac{\partial}{\partial \theta} - \frac{i}{\sin \theta} \frac{\partial}{\partial \phi} \bigg) \big[(\sin \theta)^{s} \eta \big].
\end{equation}
According to the definition (\ref{SWH-definition}), we are going to derive the explicit expressions for ${}_{\pm 1} Y_{lm}$ and ${}_{\pm 2} Y_{lm}$ in this appendix. 

If we define two differential operators $\hat{S}_{\pm}$ as
\begin{eqnarray}
\hat{S}_{\pm} \equiv -\bigg( \frac{\partial}{\partial \theta} \pm \frac{i}{\sin \theta} \frac{\partial}{\partial \phi} \bigg),
\end{eqnarray}
then the action of $\eth$ (spin-raising operator) and $\bar{\eth}$ (spin-lowering operator) on $\eta$, which is a quantity of spin weight $s$, can be written as
\begin{eqnarray}
\eth \eta = \hat{S}_+ ~\eta + s \cot \theta ~ \eta, \nonumber \\
\bar{\eth} \eta = \hat{S}_+ ~\eta - s \cot \theta ~ \eta.
\end{eqnarray}
According to the definition of the spin-weighted spherical harmonics (\ref{SWH-definition}), we find for $s = \pm 1$
\begin{eqnarray}
{}_{\pm 1} Y_{lm} &=& \pm \sqrt{\frac{(l-1)!}{(l+1)!}} ~\hat{S}_{\pm} Y_{lm}
\end{eqnarray}
and for $s=\pm 2$
\begin{eqnarray}
{}_{\pm 2} Y_{lm} 
&=& \sqrt{\frac{(l-2)!}{(l+2)!}} ~\big(\hat{S}_{\pm}^2 + \cot \theta ~\hat{S}_{\pm} \big) Y_{lm} \nonumber\\
&=& \sqrt{\frac{(l-2)!}{(l+2)!}} \bigg[\partial^2_\theta -\cot \theta \partial_{\theta} \pm \frac{2i}{\sin \theta} (\partial_{\theta} - \cot \theta) \partial_{\phi} \nonumber\\
&&- \frac{1}{\sin^2\theta} \partial^2_{\phi} \bigg] Y_{lm}
\end{eqnarray}
since $Y_{lm}$ is of spin-weight 0.

Taking into account the definition of the spherical harmonics,
\begin{equation*}\label{SH}
Y_{lm}(\theta, \phi) = (-1)^m\sqrt{\frac{(2l+1)}{4\pi} \frac{(l-m)!}{(l+m)!}} P_l^m(\cos \theta) e^{i m \phi},
\end{equation*}
and the recurrence relation for the associated Legendre function
\begin{eqnarray*}
\partial_{\theta} P_l^m &=& \frac{1}{2} \big[ P_l^{m+1} - (l+m)(l-m+1) P_l^{m-1} \big],
\end{eqnarray*}
we obtain the explicit expressions for the $s = \pm 1, \pm 2$ spin-weighted spherical harmonics in terms of Legendre function,
\begin{widetext}
\begin{eqnarray}\label{SWH1}
{}_{\pm 1} Y_{lm}
=  \frac{1}{2} (-1)^m \sqrt{\frac{2l+1}{4\pi} \frac{(l-1)!}{(l+1)!} \frac{(l-m)!}{(l+m)!}} \bigg(\mp P_l^{m+1} \pm c^2_{lm} P_l^{m-1} 
+ \frac{2m}{\sin \theta} P_l^m \bigg) e^{i m\phi},
\end{eqnarray}
\begin{eqnarray}\label{SWH2}
{}_{\pm 2} Y_{lm} &=& \frac{1}{4} (-1)^m \sqrt{\frac{2l+1}{4\pi}  \frac{(l-2)!}{(l+2)!} \frac{(l-m)!}{(l+m)!}} \bigg[ P_l^{m+2} + c^2_{lm} c^2_{l, m-1} P_l^{m-2}
- 2 \bigg(\cot \theta \pm \frac{2m}{\sin \theta} \bigg) P_l^{m+1}  \nonumber\\
&& + 2\bigg(\cot \theta \pm \frac{2m}{\sin \theta} \bigg) c^2_{lm} P_l^{m-1}
+\bigg(-c^2_{lm} - c^2_{l, m+1} + \frac{4m^2}{\sin^2 \theta} \pm \frac{8m \cot \theta}{\sin \theta} \bigg) P_l^m \bigg] e^{i m \phi},
\end{eqnarray}
\end{widetext}
where $c_{lm} = \sqrt{(l+m)(l-m+1)}$. 

\section{Angular integrals \label{appendix:angular12}}
Eqs. (\ref{SWH1}) and (\ref{SWH2}) can be further transformed to eliminate the 
trigonometric functions by using the recurrence properties of the associated Legendre functions \cite{Abramowitz}. Finally, the calculation of the 
angular integrals is reduced to the evaluation of the integrals of the products of three associated Legendre functions, 
\begin{equation}\label{IP3}
\mathcal{I}_{P3}(l_1, m_1; l_2, m_2; l_3, m_3) = \int_{-1}^1 \mathrm d \mu ~ P_{l_1}^{m_1} P_{l_2}^{m_2} P_{l_3}^{m_3},
\end{equation}
where $\mu=\cos \theta$.
The integral $\mathcal{I}_{P3}$ has an analytic result and is shown in  Appendix \ref{appendix:3LF}. Therefore, the angular integrals can always
be calculated analytically although the calculation is tedious in the general case. 

For simplicity we consider a special case where the angular momentum of the QNMs along the $z$ direction is zero, that is, $m_0=0$. 
This simplifies the calculation a lot and is sufficient to demonstrate quantum particle generation by the QNMs.
It is easy to show that, if we let $m=0$ in Eqs. (\ref{SWH1}) and (\ref{SWH2}),
\begin{eqnarray}
{}_{\pm 1} Y_{l0} &=& \mp  \sqrt{\frac{2l+1}{4\pi} \frac{(l-1)!}{(l+1)!}} ~P_l^1, \nonumber\\
{}_{\pm 2} Y_{l0} &=& \sqrt{\frac{2l+1}{4\pi} \frac{(l-2)!}{(l+2)!}} ~P_l^2.
\end{eqnarray}
From the definitions of the angular integrals, we immediately find  
\begin{eqnarray}
\mathcal{I}_{a1} &=& \sqrt{2} \int \mathrm d \Omega ~ {}_{-1}Y_{l_0 0} 
Y^*_{lm} \bigg(\frac{i}{\sin \theta}\partial_{\phi} Y^*_{l'm'} \bigg), \\ 
\nonumber\\
\mathcal{I}_{a2} &=& \int \mathrm d \Omega ~ {}_{-2}Y_{l_0 0} 
\bigg[ (\partial_{\theta} Y^*_{lm}) \bigg(\frac{i}{\sin \theta}\partial_{\phi} Y^*_{l'm'} \bigg) \nonumber\\
&& + \bigg(\frac{i}{\sin \theta}\partial_{\phi} Y^*_{lm} \bigg) (\partial_{\theta} Y^*_{l'm'}) \bigg]. \\
\nonumber\\
\mathcal{I}_{a4} &=& \sqrt{2} \int \mathrm d \Omega ~ {}_{-1}Y_{l_0 0} 
Y^*_{lm} \big(\partial_{\theta} Y^*_{l'm'} \big), \\ 
\nonumber\\
\mathcal{I}_{a5} &=& \int \mathrm d \Omega ~ {}_{-2}Y_{l_0 0} 
\bigg[ (\partial_{\theta} Y^*_{lm}) (\partial_{\theta} Y^*_{l'm'})  \nonumber\\
&& + \bigg(\frac{i}{\sin \theta}\partial_{\phi} Y^*_{lm} \bigg) \bigg(\frac{i}{\sin \theta} \partial_{\phi} Y^*_{l'm'} \bigg) \bigg]. 
\end{eqnarray}
The integration over $\phi$ gives rise to a $\delta$-function $\delta_{m', -m}$, which implies that the producing pair of 
particles have opposite angular momentum along the $z$ direction. This is not surprising given that $m_0=0$ and is 
an indication of the angular momentum conservation in the particle production process. 

Using another recurrence relation
\begin{eqnarray}
\frac{m}{\sin \theta} P_l^m  &=& -\frac{1}{2} \big[ P_{l-1}^{m+1} + (l+m)(l+m-1) P_{l-1}^{m-1} \big], \nonumber
\end{eqnarray}
of the associated Legendre function, these angular integrals become
\begin{widetext}
\begin{eqnarray}\label{AngularIntegral:1}
\mathcal{I}_{a1} &=& -\sqrt{2} \pi \delta_{m', -m} ~ \mathcal K^{lml'm'}_{l_0 1}
 \int_{-1}^{1} \mathrm d \mu ~ P_{l_0}^1 P_l^m \big[ P_{l'-1}^{m'+1} + (l'+m')(l'+m'-1) P_{l'-1}^{m'-1} \big], 
\end{eqnarray}
\begin{eqnarray}\label{AngularIntegral:2}
\mathcal{I}_{a2} &=& -\frac{\pi}{2} \delta_{m', -m} ~ \mathcal K^{lml'm'}_{l_0 2} 
\int_{-1}^{1} \mathrm d \mu ~ P_{l_0}^2 \bigg\{ \big[ P_l^{m+1} - (l+m)(l-m+1) P_l^{m-1} \big] 
\big[ P_{l'-1}^{m'+1} + (l'+m')(l'+m'-1) P_{l'-1}^{m'-1} \big] \nonumber\\
&&+  \big[ P_{l-1}^{m+1} + (l+m)(l+m-1) P_{l-1}^{m-1} \big] \big[ P_{l'}^{m'+1} - (l'+m')(l'-m'+1) P_{l'}^{m'-1} \big] 
\bigg\},
\end{eqnarray}
\begin{eqnarray}\label{AngularIntegral:4}
\mathcal{I}_{a4} &=& \sqrt{2} \pi \delta_{m', -m} ~ \mathcal K^{lml'm'}_{l_0 1}
\int_{-1}^{1} \mathrm d \mu ~ P_{l_0}^1 P_l^m \big[ P_{l'}^{m'+1} - (l'+m')(l'-m'+1) P_{l'}^{m'-1} \big],  
\end{eqnarray}
\begin{eqnarray}\label{AngularIntegral:5}
\mathcal{I}_{a5} &=& \frac{\pi}{2} \delta_{m', -m}~ \mathcal K^{lml'm'}_{l_0 2}
\int_{-1}^{1} \mathrm d \mu ~ P_{l_0}^2 \bigg\{  \big[ P_l^{m+1} - (l+m)(l-m+1) P_l^{m-1} \big] 
\big[ P_{l'}^{m'+1} - (l'+m')(l'-m'+1) P_{l'}^{m'-1} \big]  \nonumber \\
&&+ \big[ P_{l-1}^{m+1} + (l+m)(l+m-1) P_{l-1}^{m-1} \big] \big[ P_{l'-1}^{m'+1} + (l'+m')(l'+m'-1) P_{l'-1}^{m'-1} \big] \bigg\}.
\end{eqnarray}
where the factor $\mathcal K^{lml'm'}_{LM}$ is defined as
\begin{eqnarray}
\mathcal K^{lml'm'}_{LM} = \sqrt{\frac{(2L+1)(2l+1)(2l'+1)}{(4\pi)^3}  \frac{(L-M)!}{(L+M)!}\frac{(l-m)!}{(l+m)!} \frac{(l'-m')!}{(l'+m')!}}.
\end{eqnarray}
\end{widetext}

\section{Integrals of the products of three Legendre functions \label{appendix:3LF}}
We need the overlap integrals of three associated Legendre functions, Eq. (\ref{IP3}), in order to calculate the angular integrals. 
The integral $\mathcal{I}_{P3}$ was calculated by Mavromatis and Alassar \cite{Mavromatis99}, albeit with a phase error in their result. Dong and Lemus \cite{Dong02} restudied 
the problem and found an expression for $\mathcal{I}_{P3}$ with positive $m_1, m_2$ and $m_3$. Here we correct the phase error in \cite{Mavromatis99}
and give an expression that is valid for all values of $m_1, m_2$ and $m_3$. 
\begin{widetext}
\begin{eqnarray}
\mathcal{I}_{P3}(l_1, m_1; l_2, m_2; l_3, m_3) &=& \frac{(|\Delta m|)!~ (-1)^{m_1+m_2+\text{min} \{m_1+m_2, m_3 \}}}
{2^{|\Delta m| + 2} ~\Gamma(|\Delta m|) ~K_{l_1 m_1} K_{l_2 m_2} K_{l_3 m_3}} \sum_L \sum_{L'} (2L+1) (2L'+1) \nonumber\\
&&\times
\begin{pmatrix}
l_1 & l_2 & L \\ 
0 & 0  & 0 
\end{pmatrix} 
\begin{pmatrix}
l_1 & l_2 & L \\ 
m_1 & m_2  & -m_1-m_2 
\end{pmatrix}
\begin{pmatrix}
L & l_3 & L' \\ 
0 & 0  & 0 
\end{pmatrix} 
\begin{pmatrix}
L & l_3 & L' \\ 
-m_1-m_2 & m_3  & - \Delta m 
\end{pmatrix} \nonumber \\
&&\times
\frac{ \big[1+(-1)^{L'+|\Delta m|} \big] ~\Gamma(L'/2) ~\Gamma((L'-|\Delta m| +1)/2)}{K_{L', |\Delta m|} ~\Gamma((L'+|\Delta m| +2)/2 ) ~\Gamma((L'+3)/2 )}, 
\end{eqnarray}
\end{widetext}
where we have defined $\Delta m = m_3 - m_1 - m_2$, $K_{lm} = \sqrt{(l-m)!/(l+m)!}$. $\text{min} \{a, b \}$ represents the minimal value of $a$ 
and $b$, $|l_1 - l_2| \le L \le l_1 + l_2$ and $|L - l_3| \le L' \le L + l_3$.


\end{document}